\documentclass[apj]{emulateapj}
\usepackage{apjfonts}
\journalinfo{The Astrophysical Journal, in press}
\submitted{Received 2004 June 14; accepted 2004 September 26}
\shorttitle{GCs with dark matter. II}
\shortauthors{Mashchenko and Sills}

\begin{document}

\title{Globular Clusters with Dark Matter Halos. II. Evolution in Tidal Field}

\author{Sergey Mashchenko and Alison Sills}

\affil{Department of Physics and Astronomy, McMaster University,
Hamilton, ON, L8S 4M1, Canada; syam,asills@physics.mcmaster.ca}

\begin{abstract}
In this second paper in our series, we continue to test primordial scenarios of
globular cluster formation which predict that globular clusters formed in the
early universe in the potential of dark matter minihalos.  In this paper we use
high-resolution $N$-body simulations to model tidal stripping experienced by
primordial dark-matter dominated globular clusters in the static gravitational
potential of the host dwarf galaxy.  We test both cuspy Navarro-Frenk-White
(NFW) and flat-core Burkert models of dark matter halos. Our primordial globular
cluster with an NFW dark matter halo survives severe tidal stripping, and after
10 orbits is still dominated by dark matter in its outskirts.  Our cluster with
Burkert dark matter halo loses almost all its dark matter to tidal stripping,
and starts losing stars at the end of our simulations. The results of this paper
reinforce our conclusion in Paper~I that current observations of globular
clusters are consistent with the primordial picture of globular cluster
formation.
\end{abstract}


\section{INTRODUCTION}

The primordial scenario of globular cluster (GC) formation was first proposed by
\citet{pee68}. Their paper considers GCs forming in the early universe out of primordial gas
fluctuations on a Jeans mass scale. Realization that the dominant mass component
of the universe is dark matter (DM) led to the revision of this idea in
\citet{pee84}, where GCs are proposed to have formed in the potential wells of
DM minihalos in the early universe. In this revised picture GCs can be
considered as some of the earliest galaxies in the universe, with a baryons
dominated core and extended DM halo. This picture of DM-dominated GCs forming in
the early universe was later considered by many authors, including
\citet{ros88}, \citet{pad97},
\citet{cen01}, \citet{bro02}, and \citet{bea03}. 

Primordial GCs could have formed somewhere between a redshift of $\sim 30$, when
the first stars in the universe are believed to have been born inside minihalos
with total masses $10^5-10^6$~$M_\odot$ \citep{yos03}, and redshift of $\sim 6$,
when the reionization of the universe was complete \citep{bec01}. It is often
assumed that the very first stellar objects produced enough Lyman-Werner
photons (capable of dissolving H$_2$ molecules, which are the main coolant in
the pristine gas) to prohibit star formation in most of the low mass ($\lesssim
10^8$~$M_\odot$) gas-rich halos. Cosmological simulations with radiative
transfer of \citet*{ric02} suggest another possibility: in their model, numerous
mini-galaxies with masses of $\lesssim 10^8$~$M_\odot$ form inside relic
cosmological H~{\sc ii} regions. The observed positive feedback can be explained
by high non-equilibrium fraction of free electrons in defunct H~{\sc ii}
regions, which leads to enhanced H$_2$ production and hence star formation.
This is an interesting mechanism for forming GCs, as it naturally explains the
observed higher specific frequency of GCs in higher density regions of the
universe (such as around cD galaxies at the center of clusters of galaxies),
where large cosmological H~{\sc ii} regions should have formed in the early
universe.

In a widely accepted hierarchical paradigm of structure formation in the universe,
small bound objects form first, later merging to form objects of increasingly
larger size. In this picture, primordial GCs would first experience a major merger
with comparable mass halos, or be accreted by a larger halo of a dwarf galaxy
size.  In their simulations of the formation of a dwarf galaxy
in the early universe (at $z\gtrsim 10$), \citet{bro02} observed a formation of
GC-type baryonic objects inside DM minihalos, which later merged to form a
dwarf galaxy. By the end of the simulations, baryonic cores, believed to
represent proto-GCs, had apparently lost their individual DM
halos. \citet{bro02} suggested that DM was lost because of the violent
relaxation accompanying the major merger.  The lack of resolution did not allow
the authors to reach more definitive conclusion as to the fate of DM in their
proto-GCs.

In the next step, dwarf galaxies containing primordial GCs would merge to
produce objects of increasingly larger size. This process should continue at the
present time.  A well-known example is that of Sagittarius dwarf galaxy, which
is presently in the process of being disrupted by the tidal forces of Milky Way
and depositing its system of GCs into the halo of our Galaxy \citep{bel03}.

In the primordial picture of GC formation, GCs can be considered as small
nucleated dwarfs accreted at some point by larger galaxies, which lose their
extra-nuclear material (baryons and DM) due to stripping by the tidal forces of
the accreting galaxy. Such mechanism was invoked to explain the formation of a
few of the largest GCs: M51 \citep{lay00}, $\omega$ Centauri \citep{tsu03,miz03}, and G1
\citep{bek04}, which is the satellite of the M31 galaxy.
In the quoted papers, these GCs are considered to be exceptions.  We suggest
instead that this method of forming these GCs is a rule rather than
exception, and is fully consistent with primordial GC formation scenarios. In
this picture, the above largest GCs are examples of the most recent GC-producing
accretion events, and represent a small fraction of the total number of
proto-GCs accreted directly by a large galaxy (Milky Way or M31) in the modern
universe (thus skipping the intermediate step of being accreted by a larger
dwarf galaxy in the early universe).

\citet{moo96} used numerical simulations of a DM-dominated GC orbiting inside the
static potential of Milky Way halo and observations of tidal tails of M2 of
\citet{gri95} to claim that primordial scenarios of GC formation are inconsistent
with observations of Galactic GCs. As discussed by \citet{bro02} and
\citet{bea03}, the model of \citeauthor{moo96} was too simplistic to address
this issue. Most importantly, the author considered all Galactic GCs being
directly accreted by the Milky Way, which is definitely not the case in the
hierarchical picture of structure formation in the universe. Moreover, the DM
subhalo model of \citet{moo96} is a lowered isothermal sphere with arbitrary
(not cosmologically normalized) structural parameters, which makes it very
different from the DM halos expected to host a GC in the early universe.

One also has to be careful with claims on the existence of ``extratidal
features'' around Galactic satellites, which includes both GCs and dwarf
spheroidal galaxies. In the case of $\omega$~Centauri, \citet{law03} showed that
the apparent ``tidal tails'' observed by \citet{leo00} around this GC are most
probably caused by spatially variable foreground dust extinction in its
outskirts. In the case of the Draco dwarf spheroidal galaxy, the detection of
apparent ``extratidal features'' around this galaxy by \citet{irw95} was later
refuted by observations of \citet{ode01} which had much higher sensitivity.  The
``extratidal'' extension of the radial surface brightness profiles of Galactic
GCs beyond a ``tidal'' King radius is almost always seen at the levels lower
than the inferred background contamination level, and could be an artifact of
the background correction procedure. One beautiful example of a GC with obvious
tidal tails is that of Palomar~5 \citep{ode03}. As we will show in this paper, a
presence of tidal tails in some GCs is not inconsistent with these GCs having
formed with DM halos in the early universe, and hence with primordial scenarios
of GC formation.

In this series of two papers, we are using high-resolution $N$-body simulations
of GCs with DM halos to test primordial scenarios of GC formation. In
particular, we are trying to answer the following questions: 1) Are there
obvious signatures of DM presence in GCs? 2) Will DM in a primordial GC survive
hierarchical process of assembling galaxies in the early universe? In the first
paper, \citet[hereafter Paper~I]{mas04b}, we considered the initial relaxation
of a stellar cluster at the center of a DM minihalo in the early universe
(around $z=7$). The structural parameters of DM halos were fixed by results of
cosmological $\Lambda$CDM simulations. The initial non-equilibrium configuration
of stellar clusters was that of \citet[hereafter MS04]{mas04a}, where we showed that purely
stellar (no DM) homogeneous isothermal spheres with the universal values of the
initial density $\rho_{i,*}=14$~$M_\odot$~pc$^{-3}$ and velocity dispersion
$\sigma_{i,*}= 1.91$~km~s$^{-1}$ collapse to form GC-like clusters, with all
bivariate correlations between structural and dynamic parameters of Galactic GCs
being accurately reproduced. We showed in Paper~I that many observational
features of Galactic GCs, used traditionally to argue that these systems are
tidally limited purely stellar clusters without DM, can be produced by the
presence of significant amounts of DM in their outskirts. In particular, in warm
collapse models (with the initial virial parameter for the stellar core of
$0.27\lesssim\nu\lesssim 1.7$) we observed the formation of an apparent
``tidal'' cutoff in surface brightness profiles, with no unusual features in
the outer parts of the velocity dispersion profile. In cold collapse models
(with $\nu\lesssim 0.27$), on the contrary, an apparent ``break'' in the outer
parts of both surface brightness and velocity dispersion profiles is seen, which
can be mistakingly interpreted as presence of ``extratidal'' stars heated by the
tidal field of the host galaxy.

The results of Paper~I are directly applicable to dynamically young
intergalactic GCs and GCs which have not experienced significant tidal stripping
in the potential of the host galaxies. In this second paper, we test the regime
of severe tidal stripping of our primordial DM-dominated GCs in the potential of
the host dwarf galaxy. As we discussed above, being accreted by a dwarf galaxy
would be a typical fate of primordial GCs formed at high redshift. We use warm
collapse models of primordial GCs from Paper~I to set up the initial conditions
for the present paper. We consider both simulations suggested \citet[hereafter
NFW]{NFW97} and observationally motivated \citet{bur95} density profiles for our
DM halos.

This paper is organized as follows. In Section~\ref{model} we describe our
models.  In Section~\ref{results} we discuss the results of the simulations. We
conclude with Section~\ref{conclusions}.

\section{MODEL}
\label{model}

\subsection{Physical Parameters of the Models}
\label{physical}

In our model, we explore the impact of tidal fields on DM-dominated GCs orbiting
inside a static potential of host dwarf galaxy with a virial mass of
$10^9$~$M_\odot$. GCs are assumed to be accreted by the dwarf galaxy at a
redshift of $z=3$. Structural parameters (virial radius $R_{\rm vir}$, scale
radius $R_s$, and concentration $C\equiv R_{\rm vir}/R_s$) of the host galaxy
are fixed by the results of cosmological $\Lambda$CDM simulations, and at $z=3$
are equal to $R_{\rm vir}=8.16$~kpc, $R_s=1.21$~kpc, and $C=6.75$ (see
eqs.~[1-9] from Paper~I). Throughout this paper we assume the following values
of the cosmological parameters: $\Omega_m=0.27$ and $H=71$~km~s$^{-1}$~Mpc$^{-1}$
\citep{spe03}. 
Galaxies with $M_{\rm vir}=10^9$~$M_\odot$ correspond to $\sim 1\sigma$ density
fluctuations collapsing at $z=3$ \citep[their Fig.~6]{bar01} and hence are very
common at that redshift.

The initial parameters of our GCs are identical to those of our warm collapse
models from Paper~I. We could not use the cold collapse model of Paper~I because
it would be an impossible task with the present day technology with the method
we are using ($N$-body simulations). Indeed, all the models presented in the
present paper for the warm collapse only case required 3.6 CPU-years to run. The
cold collapse models from Paper~I (models C$_{n,b}$) took $\sim 7$ times longer
to run than the warm collapse ones (models W$_{n,b}$) for the same evolution
time, which is mainly due to $\sim 6$ times shorter stellar crossing time in the
cold models. In addition, to run the cold collapse models of Paper~I for the
much longer evolution time of the present paper (3~Gyr), we would have to
increase significantly the accuracy of the integration, and increase the number
of DM particles by a factor of 20 (up to $10^7$ particles) to avoid artificial
mass segregation effects which should be very significant for such long
evolution time.  As a result, the time required to simulate the cold collapse
models for 3~Gyr would be two or more orders of magnitude longer than for the
warm collapse case, or a few hundred CPU-years, which is obviously not
feasible. The reasons for choosing the warm versus hot collapse model are that
the warm one corresponds to a more typical GC (with the mass of $\sim
10^5$~M$_\odot$ for the warm model versus $\sim 10^4$~M$_\odot$ for the hot
one), and that the warm model of Paper~I exhibits interesting tidal-like
features in the outskirts of the cluster (namely, a cutoff in the surface
brightness profile) caused by the presence of DM halo, whereas the hot collapse
model does not have any such features.

An isothermal homogeneous stellar sphere with the universal
values of the initial stellar density and velocity dispersion, $\rho_{i,*}=
14$~$M_\odot$~pc$^{-3}$ and $\sigma_{i,*}= 1.91$~km~s$^{-1}$ (MS04), is set at
the center of a DM halo with either NFW or Burkert density profiles. The virial
mass of the DM halo is $m_{\rm DM}=10^7$~$M_\odot$. The stellar mass is
$m_*=8.8\times 10^4$~$M_\odot$, so the baryonic-to-DM mass ratio is
$\chi=0.0088$. Our fiducial value of $\chi$, on one hand, is larger than the
fraction of baryons in GCs of $\simeq 0.0025$ \citep{mcl99}, and on the other
hand, is smaller than the universal baryonic-to-DM density ratio of $\simeq 0.20$
\citep{spe03}. We assumed GCs have formed at $z=7$, so the structural
parameters of their DM halos are $r_{\rm vir}=885$~pc, $r_s=181$~pc, and
$c=4.88$ (from eqs.~[1-9] of Paper~I).

The initial virial ratio for our stellar clusters is $\nu_*=0.54$.  The initial
stellar radius is 11.2~pc. As we showed in Paper~I, in the absence of DM our
stellar cluster will first collapse (with the smallest value of the half-mass
radius of $r_{*,\rm min}=3.0$~pc), and then bounce to form a relaxed cluster
with a flat core, with the radius of $r_0=3.0$~pc, and a steep power-law outer
density profile. The crossing time at the half-mass radius of the relaxed
cluster is $\tau_*=0.49$~Myr. The crossing times at the half-mass radius for our
DM halos are 52 and 61~Myr for NFW and Burkert profiles, respectively.

For consistency, GCs with NFW halos are assumed to orbit a host galaxy with NFW
profile, and GCs with Burkert halos orbit a Burkert host galaxy.  In all models,
the pericentric distance is $R_p=R_s/2=0.60$~kpc, and the apocentric distance is
$R_a=R_s\times 2.5=3.02$~kpc. We chose such a small value of $R_p$ to explore a
regime of strong tidal stripping, potentially approaching a violent relaxation
mode of DM stripping in the GC formation simulations of \citet{bro02}.  Our
apocentric-to-pericentric distance ratio $R_a/R_p=5$ is similar to orbits of
substructure in cosmological $\Lambda$CDM simulations.  The theoretical radial
orbital period is $P=320$~Myr for the host galaxy with NFW profile, and
$P=340$~Myr for the Burkert case. In our simulations, subhalos experience
dynamic friction caused by the motion of the remnant in the halo of tidally
stripped DM particles. This effect is the most noticeable during first 1--2
orbits (for the following orbits the mass of the subhalo becomes too small to be
affected by the dynamic friction), and leads to slightly smaller values of
$R_p$, $R_a$, and $P$.  All our simulations are run for $t_2=3$~Gyr. (This
corresponds to the range of redshifts of $z=3\dots 1.2$.) The actual radial
orbital periods in our models are $P=270$~Myr for NFW potential and 290~Myr for
Burkert potential, so the total number of orbits is 10--11.

\subsection{Numerical Parameters of the Models}

\begin{table}
\caption{Numerical parameters of the models\label{tab1}} 
\begin{center}
\begin{tabular}{cccccccc}
\tableline
Model & $N_*$ & $\epsilon_*$ & $N_{\rm DM}$ & $\epsilon_{\rm DM}$ & $t_2$ &
$\Delta t_{\rm max}$&Note \\ & & pc & & pc & Gyr & Gyr & \\
\tableline
S      & $10^4$&   0.30       &       \nodata &  \nodata            &     3         &$2\times 10^{-4}$     &1 \\
D$_n$  &\nodata&  \nodata     &        $10^6$ &  1.5                &     3         &$2\times 10^{-4}$     &2 \\
D$_b$  &\nodata&  \nodata     &        $10^6$ &  1.5                &     3         &$2\times 10^{-4}$     &2 \\
SD$_n$ & $10^4$&   0.30       &        $10^6$ &  1.5                &     3         &$2\times 10^{-4}$     &3 \\
SD$_b$ & $10^4$&   0.30       &        $10^6$ &  1.5                &     3         &$2\times 10^{-4}$     &3 \\
DO$_n$ &\nodata&  \nodata     &        $10^6$ &  1.5                &     3         &$2\times 10^{-4}$     &4 \\
DO$_b$ &\nodata&  \nodata     &        $10^6$ &  1.5                &     3         &$2\times 10^{-4}$     &4 \\
SDO$_n$& $10^4$&   0.30       &        $10^6$ &  1.5                &     3         &$2\times 10^{-4}$     &5 \\
SDO$_b$& $10^4$&   0.30       &        $10^6$ &  1.5                &     3         &$2\times 10^{-4}$     &5 \\
\tableline
\end{tabular}
\end{center}
\tablecomments{
1) Stars only. 2) DM only. 3) Stars $+$ DM halo. 4) DM only halos
on orbit inside a static potential.  5) DM halos with stellar cores on orbit
inside a static potential. Here $N_*$ and $N_{\rm DM}$ are the number of stellar
and DM particles; $\epsilon_*$ and $\epsilon_{\rm DM}$ are the softening lengths
for stars and DM; $t_2$ and $\Delta t_{\rm max}$ are the total evolution time
and the maximum value for individual timesteps.
}
\end{table}

In total, we run 9 different simulations (see Table~\ref{tab1}). Models S (stars
only), D$_{n,b}$ (DM only), and SD$_{n,b}$ (DM $+$ stars) are evolved in
isolation (no external static gravitational field), and are used as reference
cases for the models DO$_{n,b}$ (DM only) and SDO$_{n,b}$ (DM $+$ stars) where
subhalos orbit the static potential of the host galaxy. (Here subscripts ``{\it n}''
and ``{\it b}'' denote NFW and Burkert DM profiles, respectively.)

For models containing DM, we generate initial distribution of DM particles using
the rejection method described in Paper~I. Halos are represented by $10^6$
particles with individual masses of $10$~$M_\odot$ and a softening length of
$\epsilon_{\rm DM}=1.5$~pc. \citet*{kaz04} showed that for studies dealing with
tidal disruption of substructure one should not use so-called ``local Maxwellian
approximation'', where local velocity distribution is assumed to be multivariate
Gaussian, to set up initial conditions, as such configuration is not in
equilibrium for cuspy DM density profiles and can lead to artificially high
disruption rate for subhalos. For our models, we explicitly use phase-space
distribution functions (DFs) to assign the components of velocity vectors to
different particles, which guarantees the central part of the halo to be in
equilibrium initially. To calculate DFs, for NFW profile we use the analytical
fitting formula of
\citet{wid00}, and for Burkert profile we use our own fitting formula (Paper~I,
Appendix~A). 

Our halos are truncated at a finite radius $r_{\rm vir}$, which results in the
outer parts of the halos being not in equilibrium. As we will see in
\S~\ref{isolated}, this effect has negligible impact on the results of our simulations.

For models containing a stellar core, stars are represented by $10^4$ equal mass
particles, with individual masses of $8.8$~$M_\odot$. The softening length for
stars is $\epsilon=0.30$~pc. Stars are set up as a homogeneous isothermal sphere
located at the center of the DM halo (for models containing DM).  Stellar
particles initially have a Maxwellian distribution of velocity vectors.  As we
showed in MS04, such initial non-equilibrium configuration of a stellar cluster
leads to formation of core-halo structure, with a radial density profile
resembling that of GCs, after the initial relaxation phase. The model of MS04
also successfully reproduces all empiric bivariate correlations between
structural and dynamic parameters of Galactic GCs, given that all proto-GCs
start with the same values of the stellar density
$\rho_{i,*}=14$~$M_\odot$~pc$^{-3}$ and velocity dispersion $\sigma_{i,*}=
1.91$~km~s$^{-1}$. We use these values of $\rho_{i,*}$ and $\sigma_{i,*}$ for
setting up our models.

Our models SDO$_n$ and SDO$_b$ are first evolved in isolation (with no static
gravitational field) for 120 and 170~Myr, respectively. This allowed stars and
DM at the center of the halo to reach a state of equilibrium.

We use a parallel version of the multistepping tree code GADGET \citep*{SYW01} to
run our simulations. The values of the code parameters which control the
accuracy of simulations are the same as for our warm collapse models W$_{n,b}$
from Paper~I. In particular, we use a very conservative (small) value of the
parameter $\eta=0.0025$ controlling the values of the individual timesteps,
which are equal to $(2\eta \epsilon/a)^{1/2}$, where $a$ is the acceleration of
a particle. Also, the individual timesteps are not allowed to be larger than
$\Delta t_{\rm max}=2\times 10^{-4}$~Gyr (see Table~\ref{tab1}). As a result, we
achieved an acceptable level of accuracy in our simulations, with the total
energy change $\Delta E_{\rm tot}$ being equal to 1.8\% after 3~Gyr (or more
than 6000 crossing times at the half-mass radius) for our purely stellar model
S. For models with DM (D$_{n,b}$ and SD$_{n,b}$) this number is significantly
smaller ($\Delta E_{\rm tot}<0.07$\%), because their total energy budget is
dominated by low density parts of the DM halo. We cannot estimate $\Delta E_{\rm
tot}$ for our static field models DO$_{n,b}$ and SDO$_{n,b}$, but we expect the
integration accuracy for these models to be comparable with that of the
corresponding models without the static field (D$_{n,b}$ and SD$_{n,b}$).

For all models, we output $100-1000$ snapshots for different moments of time.
In every snapshot we identify a gravitationally bound structure (if present)
using the same routine as in Paper~I. This procedure consists of two main steps.
1) We use the program {\tt addgravity}, which is based on the algorithm of
\citet{deh00} and is part of the
NEMO\footnote{\url{http://bima.astro.umd.edu/nemo/}} software package, to assign
local gravitational potential $\Phi$ values to individual particles (both DM and
stars). We use the softening length values from Table~\ref{tab1} for purely
stellar and purely DM models, and an intermediate value of $\epsilon=1$~pc for
hybrid models.  Next, we find the 1000 (100 for model S) particles with lowest
potential. For these particles we calculate weighted six phase-space components
of the halo center, using a normalized potential $\psi_i\equiv (\Phi_{\rm
max}-\Phi_i)/(\Phi_{\rm max}-\Phi_{\rm min})$ as a weight.  Then we recenter the
snapshot both spatially and in velocity to the halo center. 2) We remove
all unbound particles (with velocity module $\upsilon>[-2G\Phi]^{1/2}$) in a
single step and recompute the potential for the remnant using {\tt addgravity}.
We repeat this unbinding procedure in an iterative manner until there are no
unbound particles.

The above procedure is reasonably fast and accurate. It failed to find a bound
subhalo only in the second half of DO$_n$ run. For a few ``failed'' snapshots we
had to use the program {\tt
SKID}\footnote{\url{http://www-hpcc.astro.washington.edu/tools/skid.html}},
which is significantly slower than our simple unbinding procedure. ({\tt SKID}
is often used to find gravitationally bound structure in the results of
cosmological simulations.)

\section{RESULTS OF SIMULATIONS}
\label{results}

\subsection{Long-Term Dynamic Evolution of the Stellar Cluster}
\label{secular}

\begin{table*}
\caption{Model parameters at the end of the simulations\label{tab2}} 
\begin{center}
\begin{tabular}{cccccccccccccccc}
\tableline
Model  &  $m_*$          & $r_{h,*}$ & $\sigma_c$&  $\rho_c$      &  $R_{h,*}$&  $R_{hb}$ &$\sigma_0$ & $\Sigma_V$      & $r_0$     & $f_0$   & $\Upsilon$             &$m_{\rm DM}$     &$r_{h,\rm DM}$&  $r_\rho$ &  $r_m$\\
       & $M_\odot$       &  pc       &km~s$^{-1}$&$M_\odot$~pc$^3$&    pc     &    pc     &km~s$^{-1}$&mag~arcsec$^{-2}$& pc        &         &$M_\odot$~$L_\odot^{-1}$& $M_\odot$       &   pc         &   pc      &  pc\\
\tableline
S\tablenotemark{a}
       &$8.80\times 10^4$&  4.73     &  4.02     &   360          &  3.65     &   2.55    & 3.82     &18.66            &  2.75     & 0.211   & 1.47                    &  \nodata        &  \nodata     &  \nodata  &  \nodata\\
S      &$8.78\times 10^4$&  4.68     &  4.05     &   1300         &  3.56     &   1.33    & 3.96     &17.91            &  1.43     & 0.105   & 1.53                    &  \nodata        &  \nodata     &  \nodata  &  \nodata\\
D$_n$  &\nodata          &  \nodata  &  \nodata  &   \nodata      &  \nodata  &   \nodata & \nodata  & \nodata         &  \nodata  & \nodata & \nodata                 &$9.68\times 10^6$&  434         &  \nodata  &  \nodata\\
D$_b$  &\nodata          &  \nodata  &  \nodata  &   \nodata      &  \nodata  &   \nodata & \nodata  & \nodata         &  \nodata  & \nodata & \nodata                 &$9.66\times 10^6$&  497         &  \nodata  &  \nodata\\
SD$_n$ &$8.80\times 10^4$&  4.31     &  4.46     &   980          &  3.26     &   1.60    & 4.41     &18.05            &  1.83     & 0.153   & 1.79                    &$9.70\times 10^6$&  424         &  7.7      &  19.1    \\
SD$_b$ &$8.80\times 10^4$&  4.77     &  4.05     &   1400         &  3.59     &   1.31    & 3.98     &17.91            &  1.42     & 0.105   & 1.56                    &$9.68\times 10^6$&  484         &  17.8     &  52.9    \\
DO$_n$ &\nodata          &  \nodata  &  \nodata  &   \nodata      &  \nodata  &   \nodata & \nodata  & \nodata         &  \nodata  & \nodata & \nodata                 &$3.83\times 10^4$&  28.4        &  \nodata  &  \nodata\\
SDO$_n$&$8.80\times 10^4$&  4.31     &  4.41     &   1200         &  3.27     &   1.44    & 4.37     &17.97            &  1.67     & 0.138   & 1.81                    &$2.67\times 10^5$&  38.0        &  8.6      &  24.5    \\
SDO$_b$&$8.77\times 10^4$&  4.72     &  4.01     &   1400         &  3.57     &   1.28    & 3.98     &17.89            &  1.40     & 0.106   & 1.56                    &$4.70\times 10^4$&  65.1        &  25.7     &  \nodata\\
\tableline
\end{tabular}
\end{center}
\tablecomments{
Here $m_*$, $r_{h,*}$, $\sigma_c$, and $\rho_c$ are mass, half-mass radius,
central velocity dispersion, and central density of bound stellar clusters;
$R_{h,*}$, $R_{hb}$, $\sigma_0$, and $\Sigma_V$ are projected half-mass radius,
half-brightness radius (where surface brightness is equal to one half of the
central surface brightness), projected central dispersion, and central surface
brightness (from eq.~[17] of Paper~I); $r_0$ and $f_0$ are the King core radius
$r_0=[9\sigma_c^2/(4\pi G \rho_c)]^{1/2}$ and the fraction of the total stellar
mass inside $r_0$; $\Upsilon$ is the apparent central mass-to-light ratio [see
eq.~(\ref{eqUpsilon})]; $m_{\rm DM}$ and $r_{h,\rm DM}$ are mass and half-mass
radius of gravitationally bound DM; $r_\rho$ is the radius where density of DM
and stars becomes equal; $r_m$ is the radius where enclosed mass of DM becomes
equal to that of stars. All parameters are averaged over $t=2.67\dots 3$~Gyr.
There is no data for the model DO$_b$ as in this model the subhalo is completely
disrupted by $t\simeq 0.79$~Gyr.  }
\tablenotetext{a}{Parameters for the moment of time $t\simeq 0.28$~Gyr.}
\end{table*} 

In the first two lines of Table~\ref{tab2} we show the model parameters for an
isolated stellar cluster without DM (model S) for two moments of time --- soon
after the initial relaxation of the cluster ($t\simeq 0.28$~Gyr; first line) and
at the end of the simulations ($t\simeq 2.84$~Gyr; second line). As you can see,
some stellar cluster parameters (such as half-mass radius $r_{h,*}$ and central
velocity dispersion $\sigma_c$) stay virtually the same throughout the
simulations, whereas others (central density $\rho_c$, central surface
brightness $\Sigma_V$, King core radius $r_0$, and core mass fraction $f_0$)
undergo significant changes.


\begin{figure}
\plotone{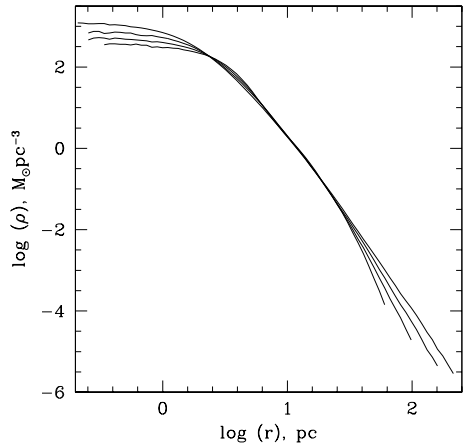}
\caption {Averaged radial stellar density profiles for model S for four time intervals:
$t=0.12\dots 0.37$, $0.37\dots 0.87$, $0.87\dots 1.61$, and $1.61\dots 3$~Gyr
(from bottom to top in the left part of the figure, and from left to right in
the right part).
\label{rho_st} }
\end{figure}


On a more detailed level, these changes can be followed in Figure~\ref{rho_st},
where we show averaged radial density profiles for model S for four different
time intervals. As you can see in this figure, the flat core of the cluster
becomes smaller and denser with time. Interestingly, the ``dent'' feature, seen
outside of the core in the radial density profile of a freshly relaxed cluster,
disappears in significantly evolved clusters. The outer part of the profile becomes
increasingly more shallow with time, with the following values of the power-law
exponent: $\gamma=-7.4$, $-5.9$, $-5.1$, and $-4.6$.

As you can see in Table~\ref{tab2}, by the end of the simulations, $\sim 50$\%
of core stars in model S evaporate. These stars stay gravitationally bound to the
cluster, populating its outer parts and making the outer density profile more
shallow.

The observed secular evolution of our model stellar cluster is very similar to
gravothermal instability (or core collapse) known to occur in real GCs. The code
we use to simulate the cluster is collisionless: it softens gravitational
potential of particles separated by less than the softening length $\epsilon_*$
(which is comparable to the average distance between particles in the cluster).
As a result, close encounters between particles are not treated
correctly. Moreover, our stellar particles do not represent individual stars, as
they are $\sim 10$ times more massive than stars in GCs. Nevertheless, it
appears that our collisionless simulations capture the essence of the long-term
dynamic evolution of GCs, probably because the main driving mechanism for such
evolution is not infrequent very close encounters between stars, but rather
the cumulative effect of numerous weak interactions \citep{S87}, which are
resolved reasonably well by our code.

The similarity between our stellar model's secular evolution and gravothermal
instability is also seen on a more quantitative level.  The idealized model of core
collapse of a GC predicts a correlation between the central density
and the radius of the core: $\rho_c\propto r_0^{-2.21}$
\citep{S87}. In our model S, we observe a very similar correlation:
$\rho_c\propto r_0^{-2.0}$ for the time interval $t=0.3\dots 3$~Gyr.  Both the
idealized theory of \citet{S87} and our model exhibit very mild evolution of
central velocity dispersion: $\sigma_c\propto r_0^{-0.10}$ and $\sigma_c\propto
r_0^{-0.01}$, respectively.

The fact that our stellar particles are $\sim 10$ times more massive than stars
in GCs results in the pace of secular evolution in our models being
significantly faster than in real GCs. To estimate how dynamically old our
models are at the end of simulations we use the analytical theory of core
collapse of \citet{S87}. The most sensitive dynamic age parameter is the core
density $\rho_c$, which according to the model of \citet{S87} evolves with time
$t$ as $\rho_c \propto (1-t/t_{\rm coll})^{-1/0.86}$, where $t_{\rm coll}$ is
the core collapse time. For our model S, $\rho_c(3{\rm
~Gyr})/\rho_c(0)=1300/360$ (see Table~\ref{tab2}), so $t/t_{\rm coll}\simeq 2/3$
at $t=3$~Gyr. Galactic GCs are known to span the whole spectrum of dynamic ages,
ranging from dynamically young with $t/t_{\rm coll}\ll 1$ (such as
$\omega$~Centauri and NGC~2419) to post-core-collapse systems with $t/t_{\rm
coll}\geqslant 1$ ($\sim 20$\% of all Galactic GCs).  As you can see, at
$t\leqslant 3$~Gyr our models span a large range of dynamic ages, corresponding
to many (probably most) Galactic GCs. In addition, the ratio of the predicted
model core collapse time $t_{\rm coll}\simeq 4.5$~Gyr to the model orbital time
$t_{\rm orb}\simeq 0.3$~Gyr is approximately the same as for dynamically old
Galactic GCs, which have $t_{\rm coll}/t_{\rm orb}\sim 15$ (assuming that
$t_{\rm coll}\sim 12$~Gyr and $t_{\rm orb}\sim 0.8$~Gyr). As a result, our
models can correctly describe both secular evolution and tidal stripping of many
Galactic GCs.

One could treat the observed long-term dynamic evolution of stars in our models
as a numeric artifact and a nuisance. Instead, by arguing that this effect
reflects main features of gravothermal instability in real GCs, we can use it to
explore the impact of long-term dynamic evolution on the properties of hybrid
(stars $+$ DM) GCs.

\subsection{Isolated Models}
\label{isolated}

In Paper~I we demonstrated that a warm collapse of a homogeneous isothermal
stellar sphere inside a live DM halo (either NFW or Burkert) produces a GC-like
cluster with an outer density cutoff in the stellar density distribution
resembling a tidal feature in King model. Our results confirmed the idea of
\citet{pee84} that a presence of DM can be an alternative explanation for
apparent ``tidal'' radial density cutoffs observed in some GCs. Here we use
models S and SD$_{n,b}$ to check if this explanation can be extended to
stellar clusters which have experienced significant secular evolution.


\begin{figure*}
\plottwo{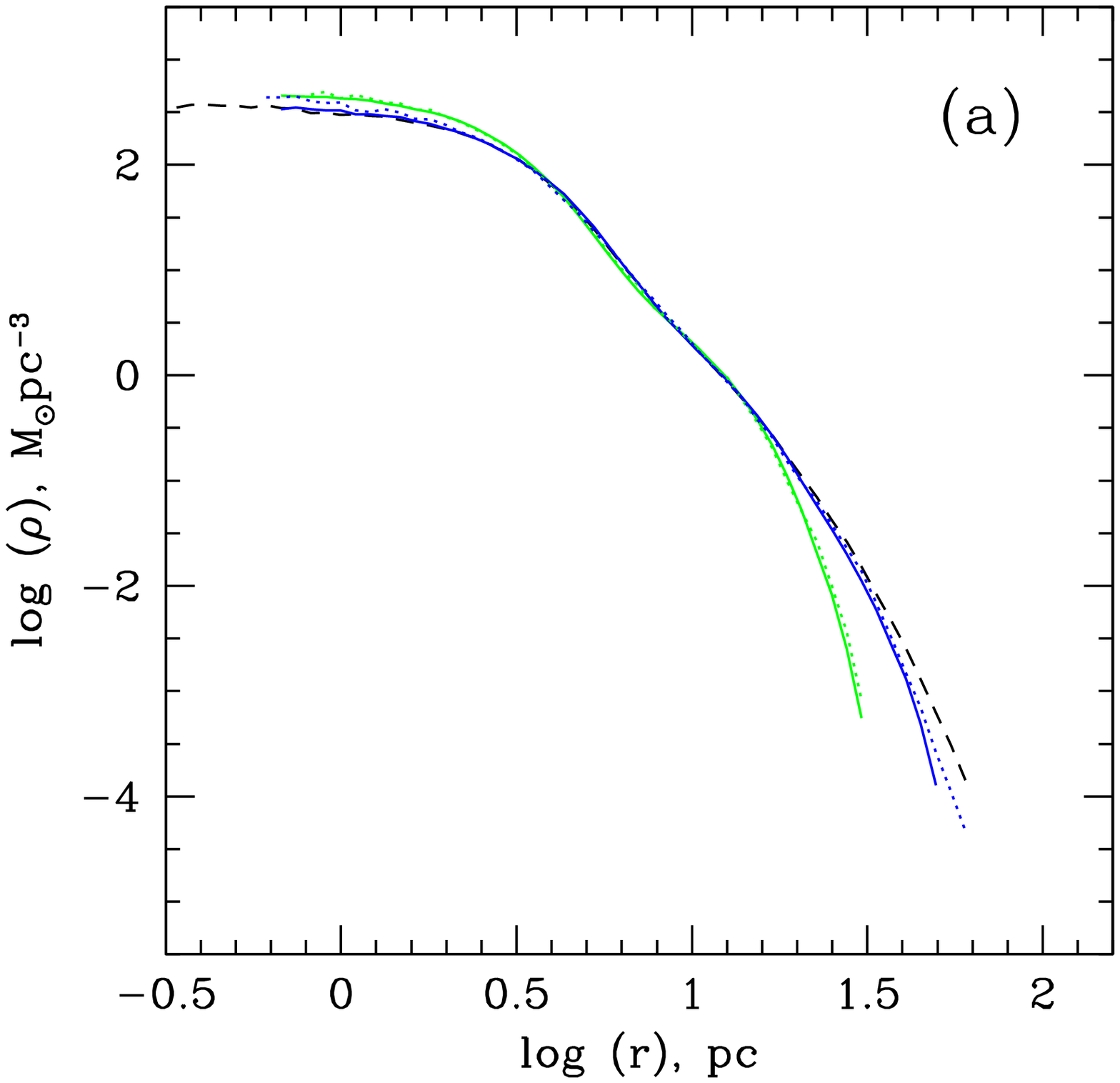}{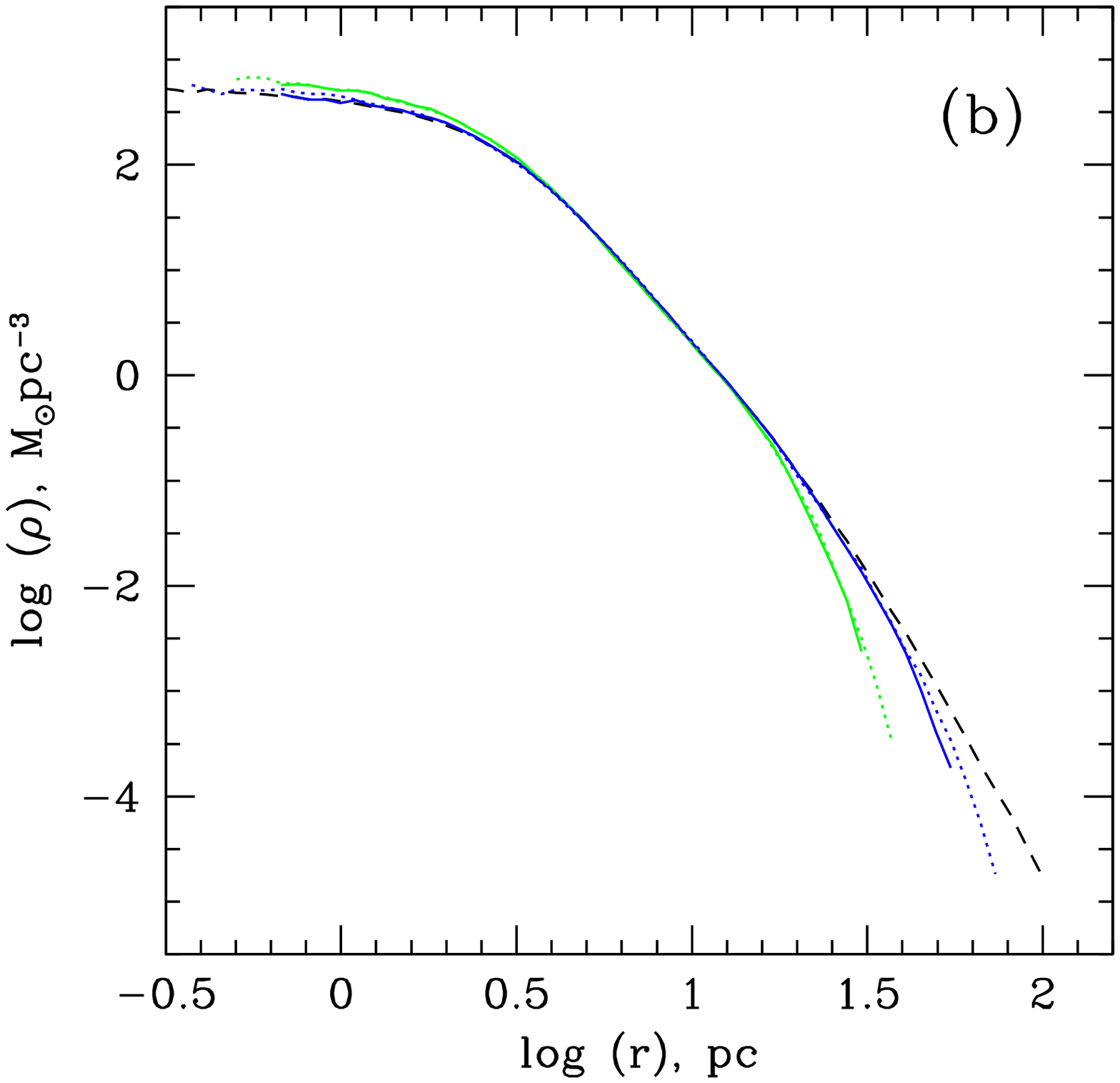}
\plottwo{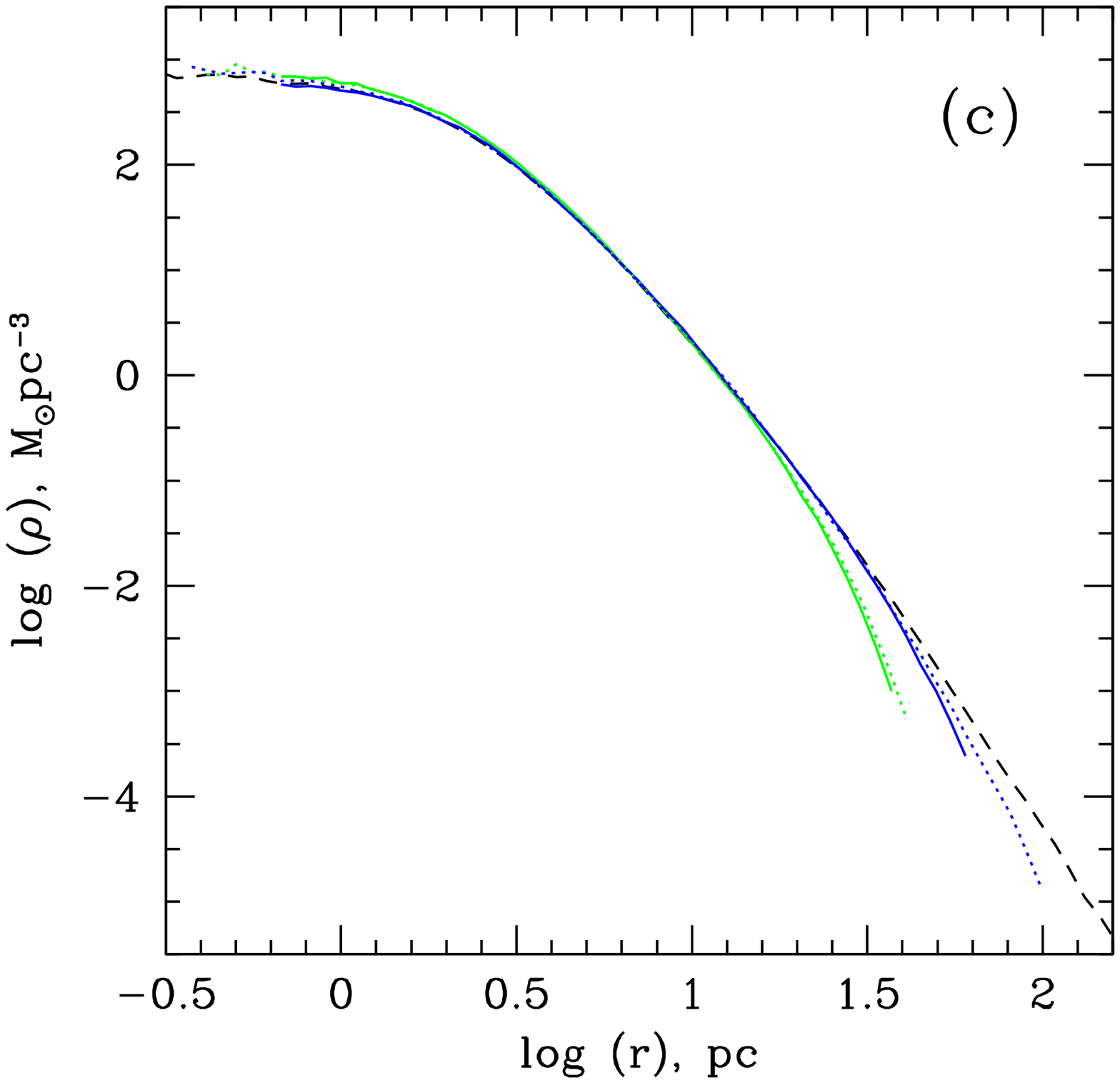}{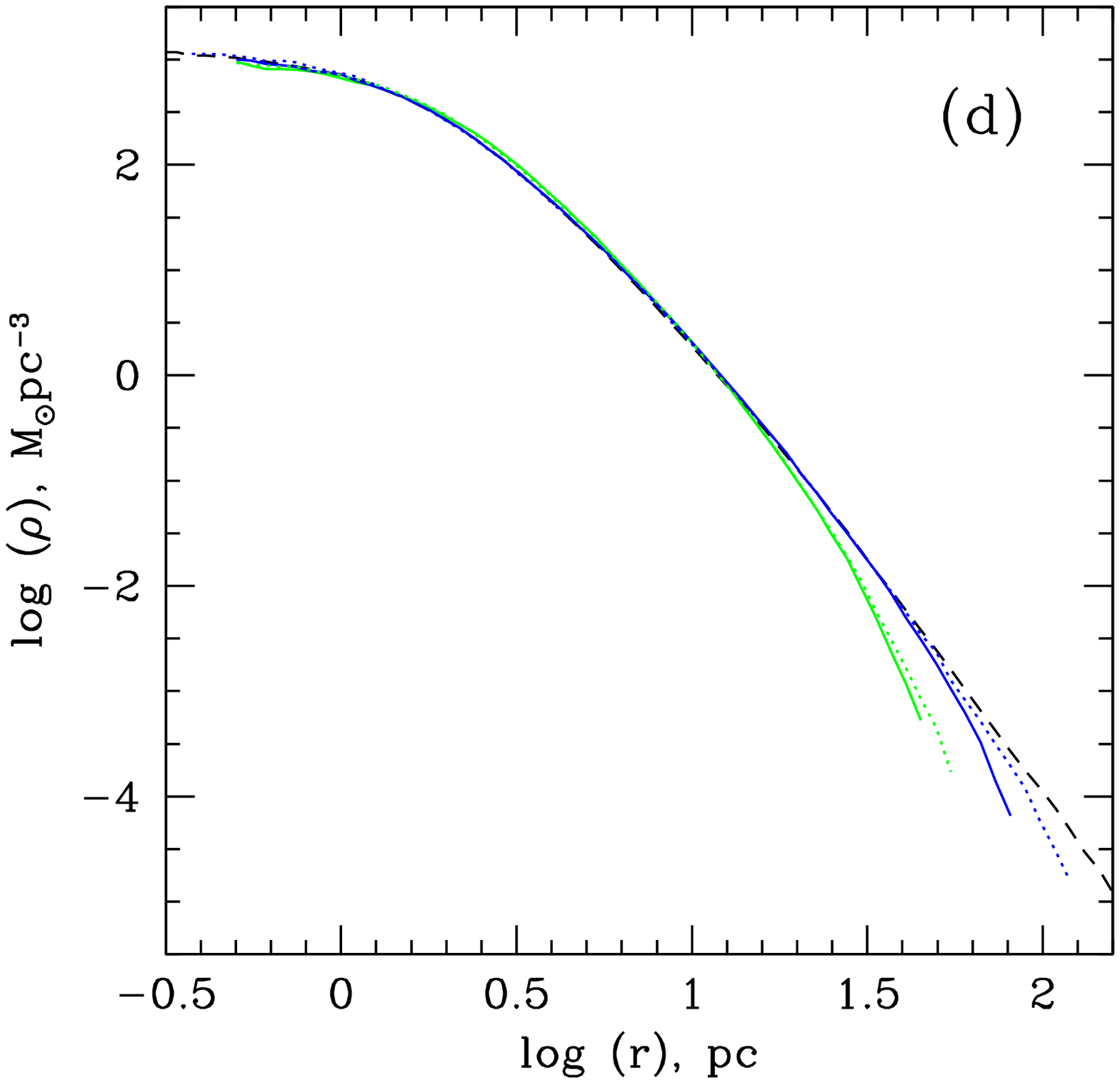}
\caption {Averaged stellar radial density profiles for four time intervals:
(a) $t=0.12\dots 0.37$~Gyr; (b) $t=0.37\dots 0.87$~Gyr; (c) $t=0.87\dots
1.61$~Gyr; (d) $t=1.61\dots 3$~Gyr (the same time intervals as in
Figure~\ref{rho_st}). In the bottom-right parts of the panels, the sequence of
models is as follows (from left to right): SD$_n$ (solid line; colored green in
electronic edition), SDO$_n$ (dotted line; colored green in electronic edition),
SD$_b$ (solid line; colored blue in electronic edition), SDO$_b$ (dotted line;
colored blue in electronic edition), and S (short-dashed line).
\label{rho} }
\end{figure*}


In Figure~\ref{rho} we show the stellar radial density profiles for models S
(short-dashed lines) and SD$_{n,b}$ (solid lines) for the same four time
intervals as in Figure~\ref{rho_st}. As you can see, flattening of the outer
density profile caused by the dynamic evolution of the cluster does not remove
an apparent cutoff feature observed in DM-dominated GCs. As time goes on, the
slope of the outer density profile becomes more shallow for both purely stellar
and hybrid GCs, but the relative change of the slope caused by the presence of
DM stays approximately the same. Also, the radius where the two profiles start
diverging, stays approximately the same ($\sim 18$~pc) for the NFW profile, and
gradually increases (from $\sim 35$ to $\sim 45$~pc) for the Burkert profile.
For both types of DM halos, this radius is very close to the radius $r_m$
where the inclosed masses of stars and DM become equal (see Table~\ref{tab2}).

The reason for the persistence of the density cutoff in the course of the
secular evolution of our cluster is the same as for the appearance of the cutoff
at the end of the initial violent relaxation phase (see Paper~I). The original
cutoff is caused by the fact that at large radii ($r\gtrsim r_m$) the potential
of the hybrid GC is dominated by DM. As a result, a smaller fraction of stars,
ejected from the cluster during the violent relaxation, can populate outer halo,
creating a density cutoff at $r\sim r_m$. Similarly, during long-term dynamic
evolution of the cluster, caused by encounters between individual stars, stars
are being ejected from the core (core evaporation), with very few of them
populating the halo beyond $r_m$ because of the potential being dominated by DM
at such large radii.

We use isolated DM-only models D$_{n,b}$ to see how numerical artifacts affect
the DM density distribution in our simulations. \citet{hay03} discussed the impact
of discreteness effects of $N$-body simulations on radial density profiles of
isotropic NFW halos. According to these authors, for the first $\sim 100$
crossing times at the virial radius, the density at the center of the halo is
decreasing due to heating by faster moving particles, causing the core to expand.
At the end of this phase, the central velocity dispersion becomes comparable to
the velocity dispersion at the scale radius $r_s$. (In NFW models the velocity
dispersion is highest around $r_s$, and becomes smaller for both smaller and
larger radii.) 


\begin{figure*}
\plottwo{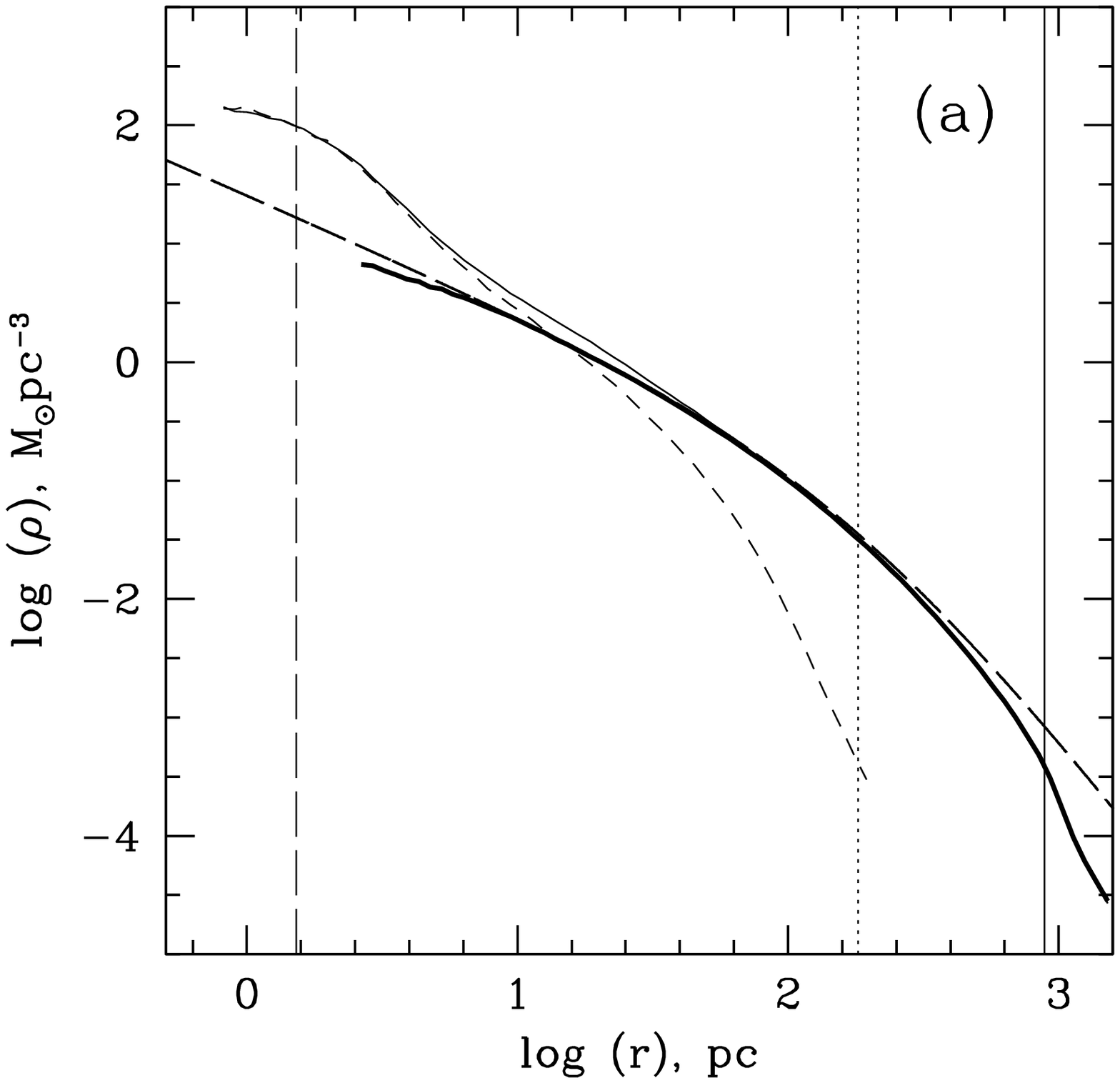}{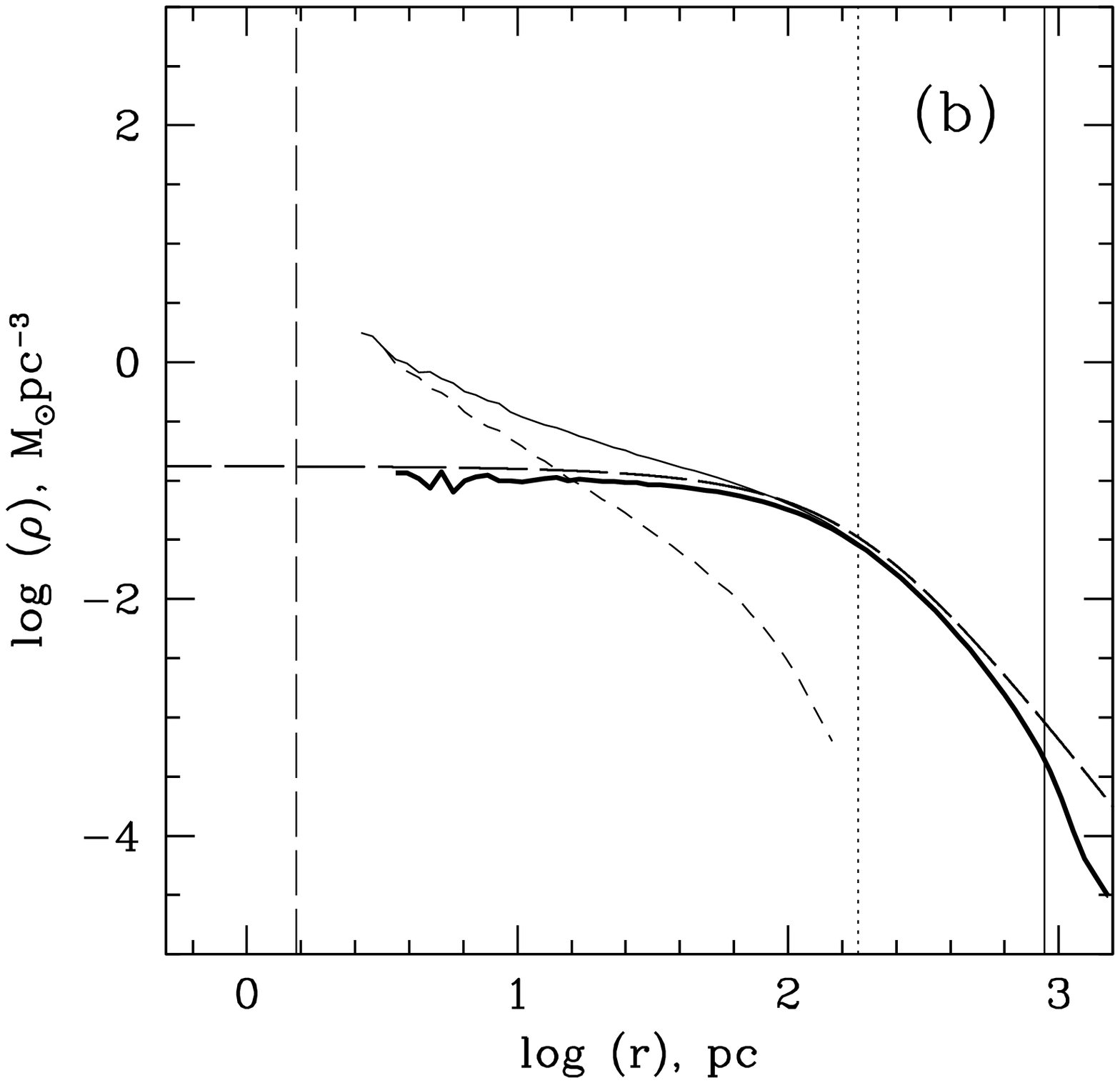}
\caption {Averaged DM radial density profiles for the time interval $t=1.61\dots 3$~Gyr
(the same time interval as in Fig.~\ref{rho}d). (a) NFW models. (b) Burkert
models.  Long-dashed lines show analytical profiles for undisturbed DM
halos. Thick solid lines correspond to models D$_{n,b}$ (no stars, no static
potential), thin solid lines show the profiles for models SD$_{n,b}$ (DM $+$
stars, no static potential), and short-dashed lines correspond to models
SDO$_{n,b}$ (DM $+$ stars on orbit in static potential). Vertical long-dashed,
dotted, and solid lines mark the values of $\epsilon_{\rm DM}$, $r_s$, and
$r_{\rm vir}$, respectively.
\label{DM_rho} }
\end{figure*}


We see a similar trend in our NFW model, D$_n$. In our case, the total evolution
time is $\sim 24$ crossing times at the virial radius $\tau_{\rm vir}=124$~Myr,
so at the end of the simulations the model is still in the core-heating
regime. As you can see in Figure~\ref{DM_rho}a, at $t\simeq 3$~Gyr the inner DM
density profile is reasonably close to the initial one down to a radius of
$\sim 2.5$~pc $\simeq 1.7 \epsilon_{\rm DM}$. 

In the case of the Burkert profile, the initial central velocity dispersion is
also smaller than the dispersion at $r_s$, though the difference is much smaller
than for the case of NFW halo. In our model D$_b$, we see a slight decrease in
the DM density in the inner part of the halo at the end of the simulations
(Figure~\ref{DM_rho}b), with a reasonably accurate profile down to a radius of
$\sim 3.5$~pc $\simeq 2.3 \epsilon_{\rm DM}$.

NFW and Burkert models would be in equilibrium only if they had infinitive size
and mass. As our models are truncated at at a finite radius $r_{\rm vir}$, the DM
density in the outer parts of the halos becomes lower with time. As you can see
in Figure~\ref{DM_rho}, even at the end of the simulations the deviation of the
outer DM density profile from the theoretical one is not significant within
$r_{\rm vir}$, and is negligible for $r\lesssim 3 r_s$. Most of the changes in
the outer density profiles happen within first one or two $\tau_{\rm vir}$, which is
comparable to the orbital period in models DO$_{n,b}$ and SDO$_{n,b}$. All the above
let us conclude that the truncation of our models at the radius of $r_{\rm vir}$
should not have a noticeable impact on the results of our simulations.

\subsection{Evolution in Tidal Field: Preliminary Analysis}
\label{prelim}

\citet{hay03} showed that an isotropic NFW subhalo, following a circular orbit
inside the static potential of a host galaxy with NFW profile, can be completely
disrupted by tidal forces after a few orbits if its tidal radius $r_{\rm tid}$
is smaller than $\sim 2 r_s$. These authors argued that the relative easiness to
disrupt an NFW satellite is linked to the fact that when an NFW halo is
instantaneously truncated at a radius $r<r_{\rm bind}\simeq 0.77 r_s$, it becomes
unbound (with the total energy $E_{\rm tot}$ of the remnant becoming
positive). The situation with singular isothermal spheres is completely
different, as such halos stay gravitationally bound for any $r$, and potentially
can survive indefinitely in an external tidal field.


\begin{figure}
\plotone{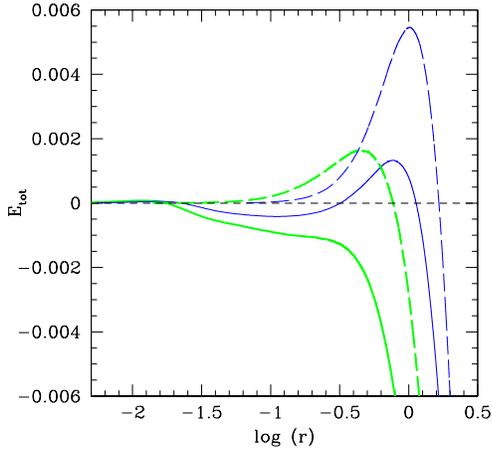}
\caption {
Total energy $E_{\rm tot}$ of halos instantaneously truncated at a radius $r$.
Here $r$ is in $r_s$ units, and $E_{\rm tot}$ is in $Gm_{\rm vir}^2/r_s$ units.
Thick and thin lines (colored green and blue in electronic edition,
respectively) correspond to NFW and Burkert profiles, respectively.  Long-dashed
lines correspond to the initial configuration of DM-only D$_{n,b}$ models, and
solid lines correspond to sufficiently relaxed ($t=0.25$~Gyr) models SD$_{n,b}$
which have a stellar core.
\label{E_tot} }
\end{figure}

\begin{figure}
\plotone{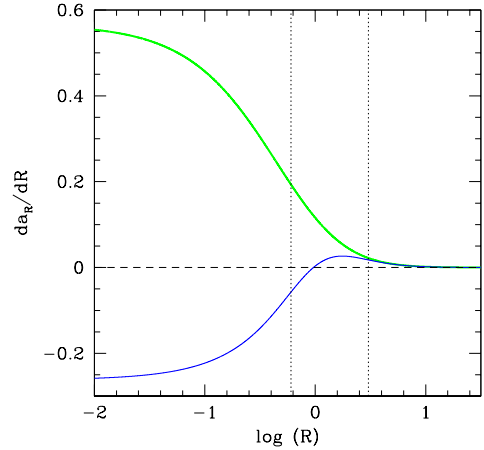}
\caption {Radial profile of tidal acceleration $da_R/dR$ for NFW (thick solid line;
colored green in electronic edition) and Burkert (thin solid line; colored blue
in electronic edition) halos with a concentration of $C=6.75$. Here $R$ is in
$R_s$ units, and $da_R/dR$ is in $GM_{\rm vir}/R_s^3$ units. Vertical dotted
lines show the pericentric and apocentric distances for the subhalo orbit in
models DO$_{n,b}$ and SDO$_{n,b}$. Horizontal short-dashed line corresponds to
zero radial tidal acceleration.
\label{a_tid} }
\end{figure}


In Figure~\ref{E_tot} we compare the binding properties of our NFW and Burkert
halos (thick and thin long-dashed lines, respectively). As you can see, Burkert
halos are similar to NFW ones in becoming unbound if truncated below a certain
radius. Burkert halos appear to be much easier to disrupt tidally than NFW
halos: in the case of Burkert profile, $r_{\rm bind}\simeq 1.66 r_s$, which is
$\sim 2.1$ times larger than for NFW profile. Also, the positive total energy of
a Burkert halo truncated to $r<r_{\rm bind}$ is $\sim 3.3$ times larger than the
corresponding quantity for an NFW halo with the same concentration and mass (see
Figure~\ref{E_tot}).

\begin{figure*}
\plottwo{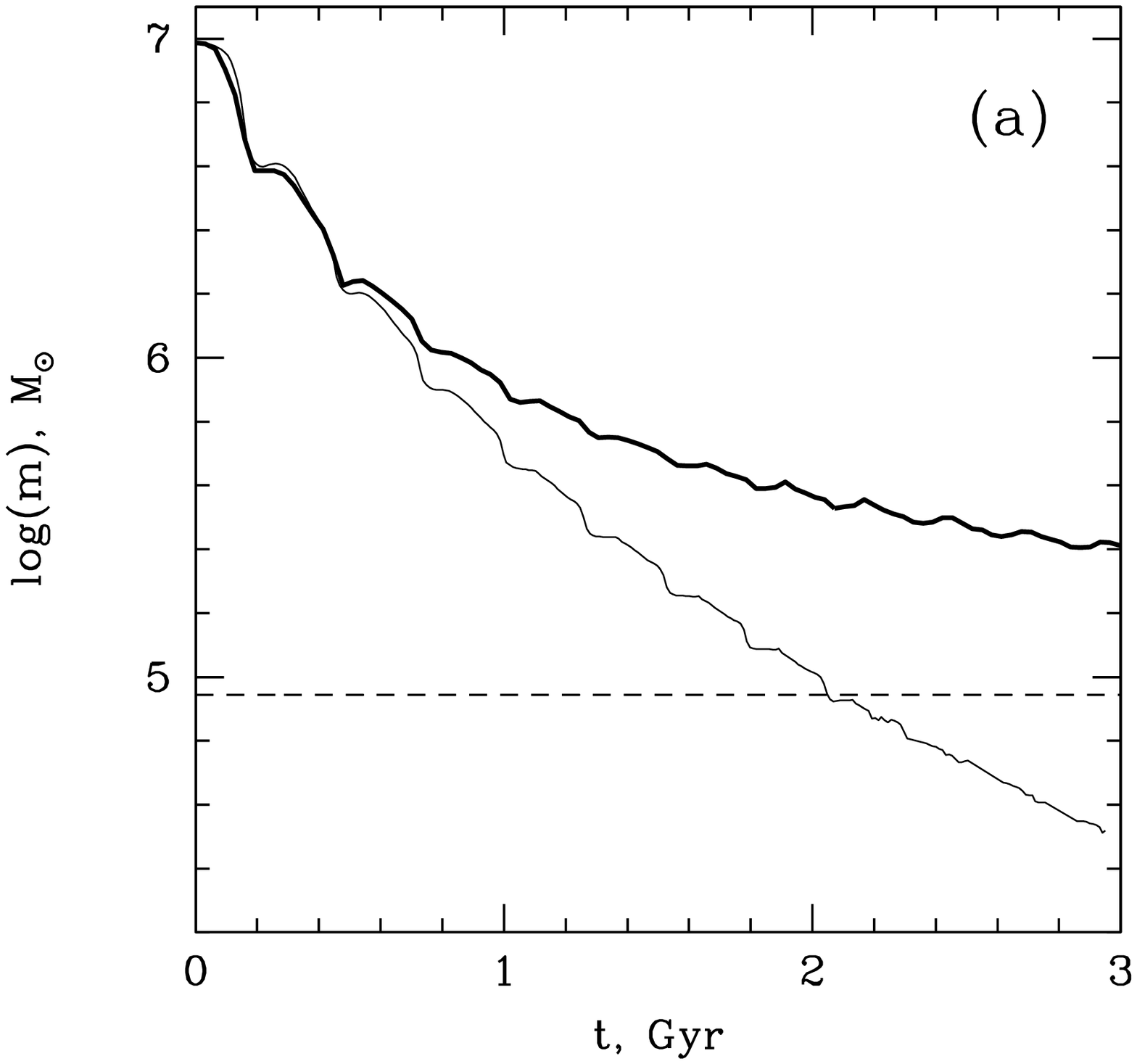}{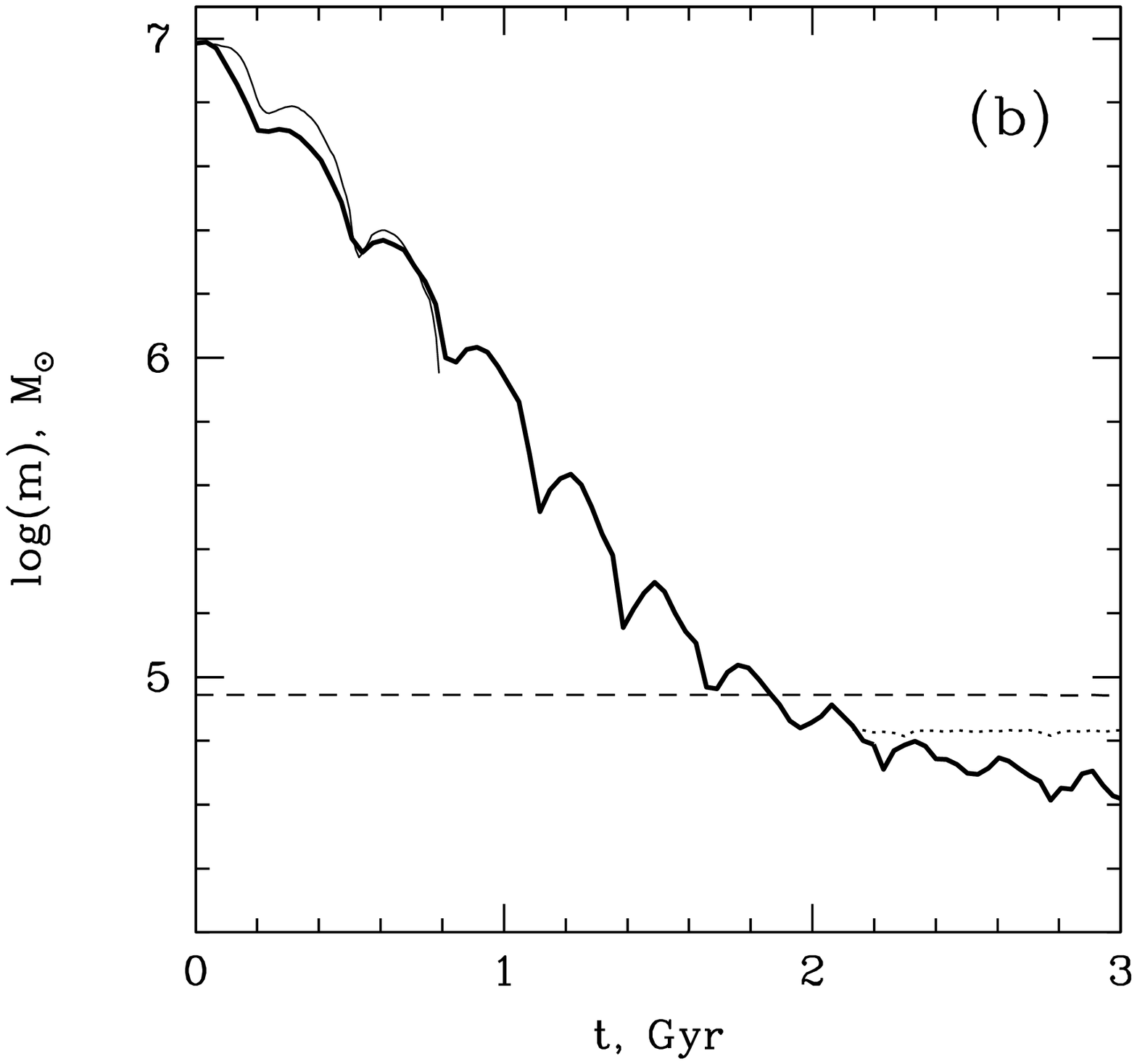}
\caption {Time evolution of the gravitationally bound DM mass of subhalos in models
SDO$_{n,b}$ (thick lines) and DO$_{n,b}$ (thin lines). (a) NFW halos (models
SDO$_n$ and DO$_n$). (b) Burkert halos (models SDO$_b$ and DO$_b$). Short-dashed
lines show the evolution of the bound stellar mass in the corresponding SDO
models. Dotted line in the panel (b) shows the evolution of DM bound mass for
subhalo from the model SDO$_b$ which at $t=2.13$~Gyr was removed from the static
potential of the host galaxy and was let to evolve in isolation.
\label{DM_mass} }
\end{figure*}

In Figure~\ref{a_tid} we compare the strength of the tidal force as a function
of radius for two types of host galaxies: with NFW profile (thick line) and with
Burkert profile (thin line). More specifically, in this figure we plot the
radial dependence of the radial gradient of gravitational acceleration 
$da_R/dR=d/dR\,[GM(R)/R^2]$. A product of this quantity on the
linear size of a subhalo gives an estimate of the differential (tidal)
acceleration between two opposite parts of the subhalo. Two vertical dotted
lines in Figure~\ref{a_tid} show the range of radial distances covered by our
models.

As you can see in Figure~\ref{a_tid}, for NFW profile the quantity 
$da_R/dR$ is always positive (meaning that the tidal force is always
stretching a subhalo in radial direction). In the small radii limit, 
$da_R/dR$ asymptotically approaches a constant, which is equal to $2/3\,
[\ln (1+C)-C/(1+C)]^{-1}\, GM_{\rm vir}/R_S^3$. For Burkert halos, the radial
tidal force reaches a maximum at $R=1.76328 R_s$, becomes equal to zero at
$R=0.96340 R_s$, and asymptotically approaches a negative constant $-2/3\,[\ln
(1+C)+1/2 \ln(1+C^2)-\arctan C]^{-1}\, GM_{\rm vir}/R_S^3$ at small radii.  The
negative sign for $da_R/dR$ means that at $R\lesssim R_s$ the radial
tidal force becomes compressing instead of being stretching. In the interval of
radii covered by the orbital motion of our subhalos, the NFW host halo has
significantly stronger radial tidal acceleration than the Burkert halo (see
Figure~\ref{a_tid}).

To summarize the above analysis, the Burkert halos are easier to disrupt tidally
than NFW halos. On the other hand, a tidal field of a Burkert host galaxy
appears to be less disrupting than that of an NFW host galaxy with the same mass and
concentration. In addition, Burkert halos have an unusual property of having a
compressing radial tidal force within the scale radius $R_s$. One has thus to
resort to numerical simulations to understand differences in substructure
evolution for NFW and Burkert cases.

\subsection{Tidal Stripping of DM-only subhalos}


\begin{figure*}
\plottwo{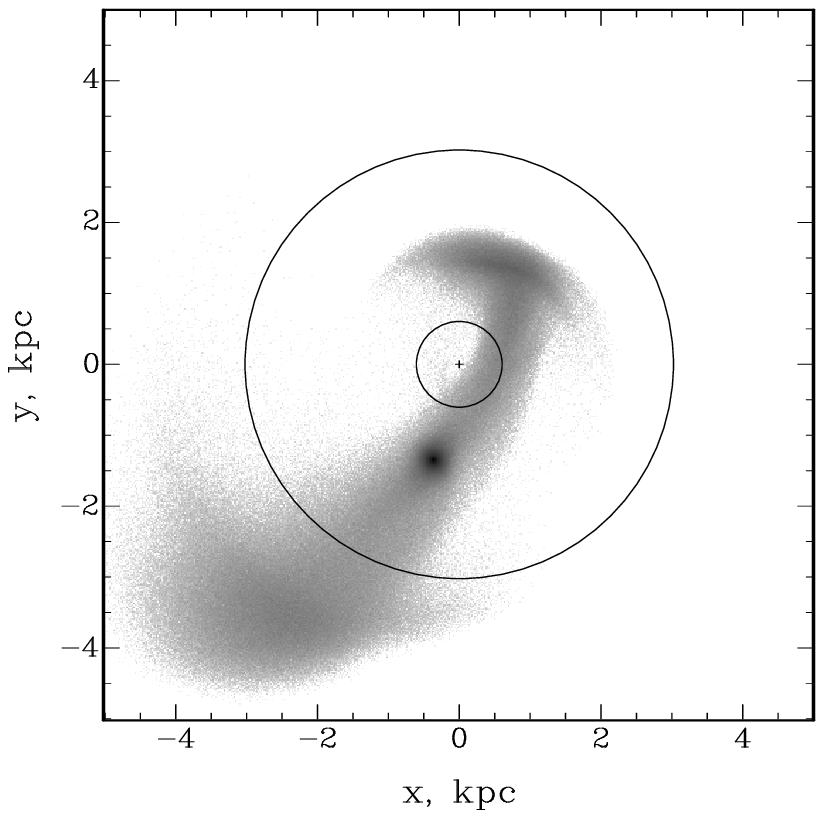}{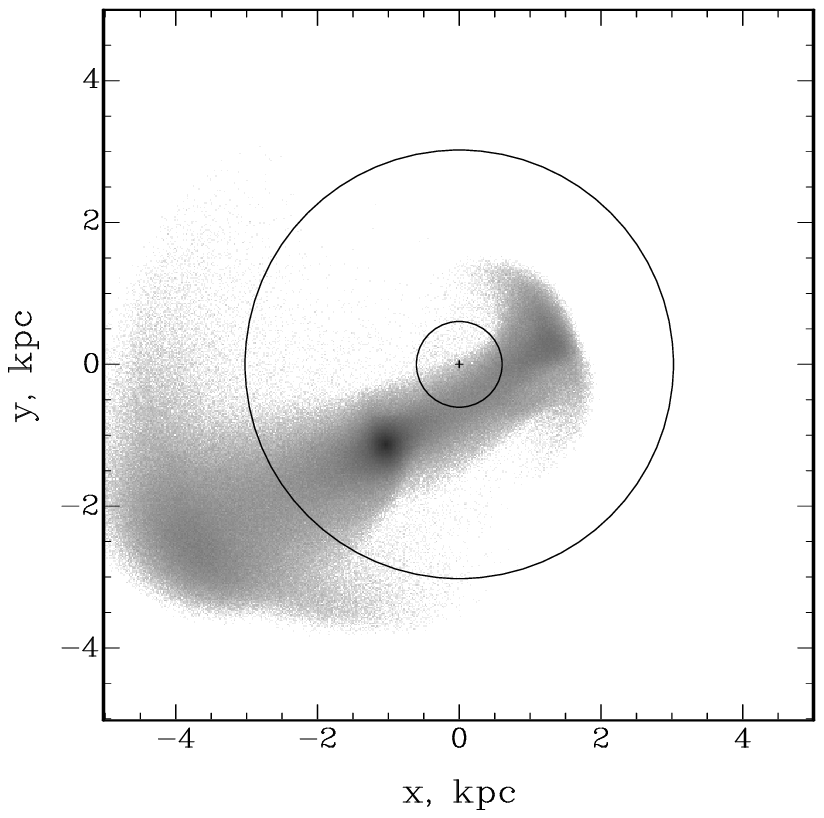}
\plottwo{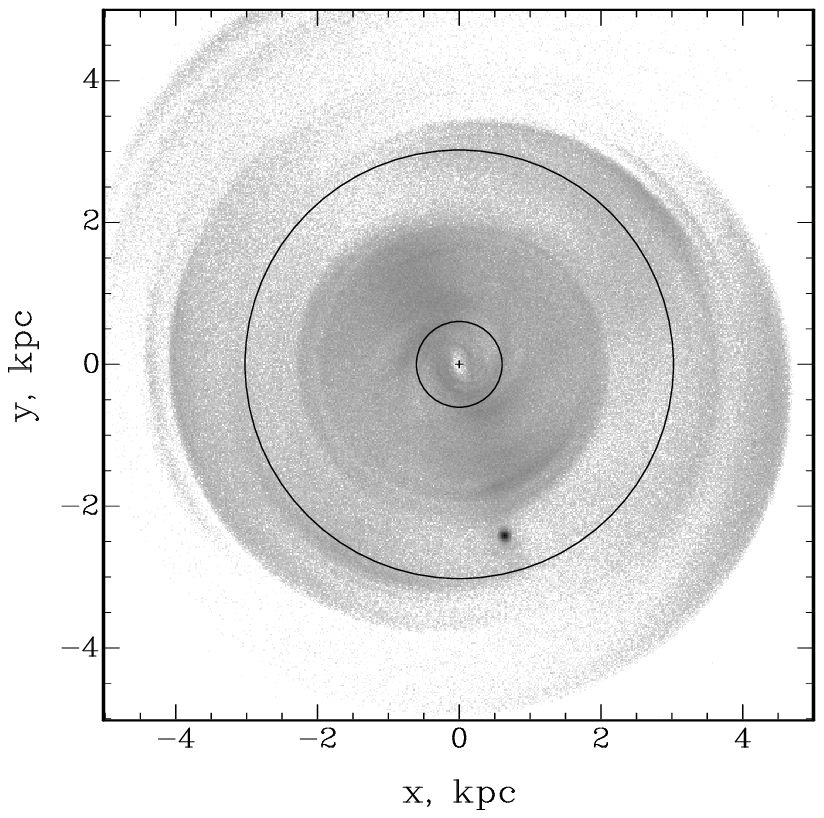}{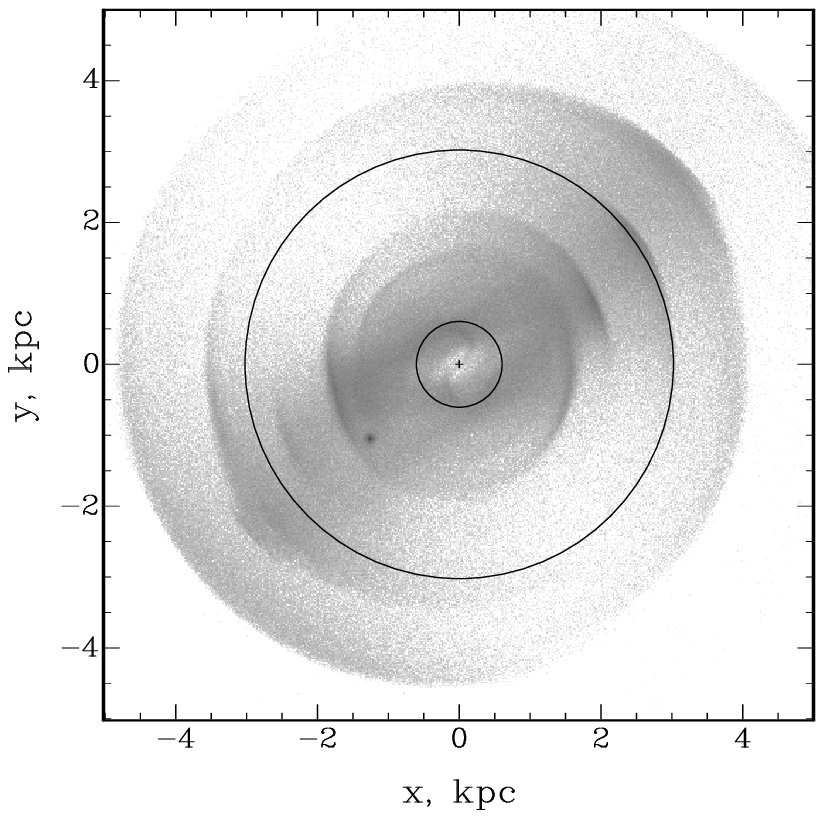}
\caption {Grey-scale maps of DM column density for models SDO$_n$ (two left panels)
and SDO$_b$ (two right panels) for two moments of time: $t\simeq 0.43$~Gyr (when
the subhalos are approaching the pericenter for the second time; two top panels)
and $t=3$~Gyr (at the end of simulations; two bottom panels).  The orbit of the
subhalo is seen face-on.  In all panels we use the same logarithmic scale for
the values of DM column density. Cross marks the location of the center of the
host galaxy, and two concentric circles correspond to apocentric and pericentric
distances of the subhalo's orbit.
\label{column} }
\end{figure*}


In our models DO$_{n,b}$, a DM-only (NFW or Burkert) subhalo is orbiting on
eccentric ($R_a/R_p=5$) orbit inside a static potential of the host (NFW or
Burkert) galaxy. We use the same definition of $r_{\rm tid}$ as \citet{hay03}:

\begin{equation}
\frac{m(r_{\rm tid})}{r_{\rm tid}^3} = \left[ 2-\frac{R}{M(R)}\frac{\partial M}{\partial R}\right]
\frac{M(R)}{R^3}.
\label{eq_rtid}
\end{equation}

\noindent Here $M(R)$ and $m(r)$ is enclosed mass as a function of radius for
the host and satellite halos, respectively. At $R=R_p$, the subhalos have the
following values of tidal radii: $r_{\rm tid}=314$~pc $=1.73 r_s$ for NFW
profile, and $r_{\rm tid}=498$~pc $=2.75 r_s$ for Burkert profile.


\begin{figure*}
\plottwo{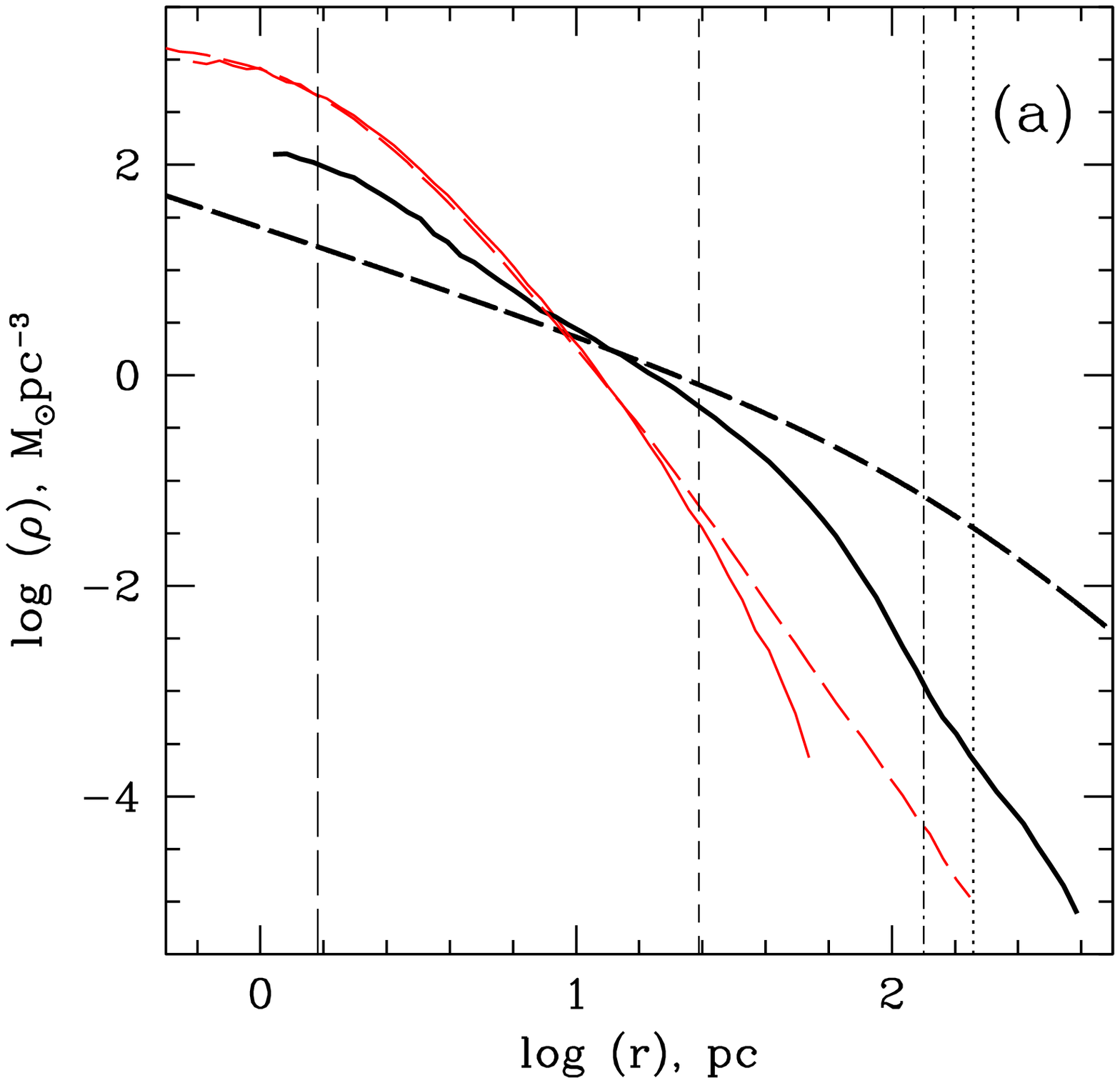}{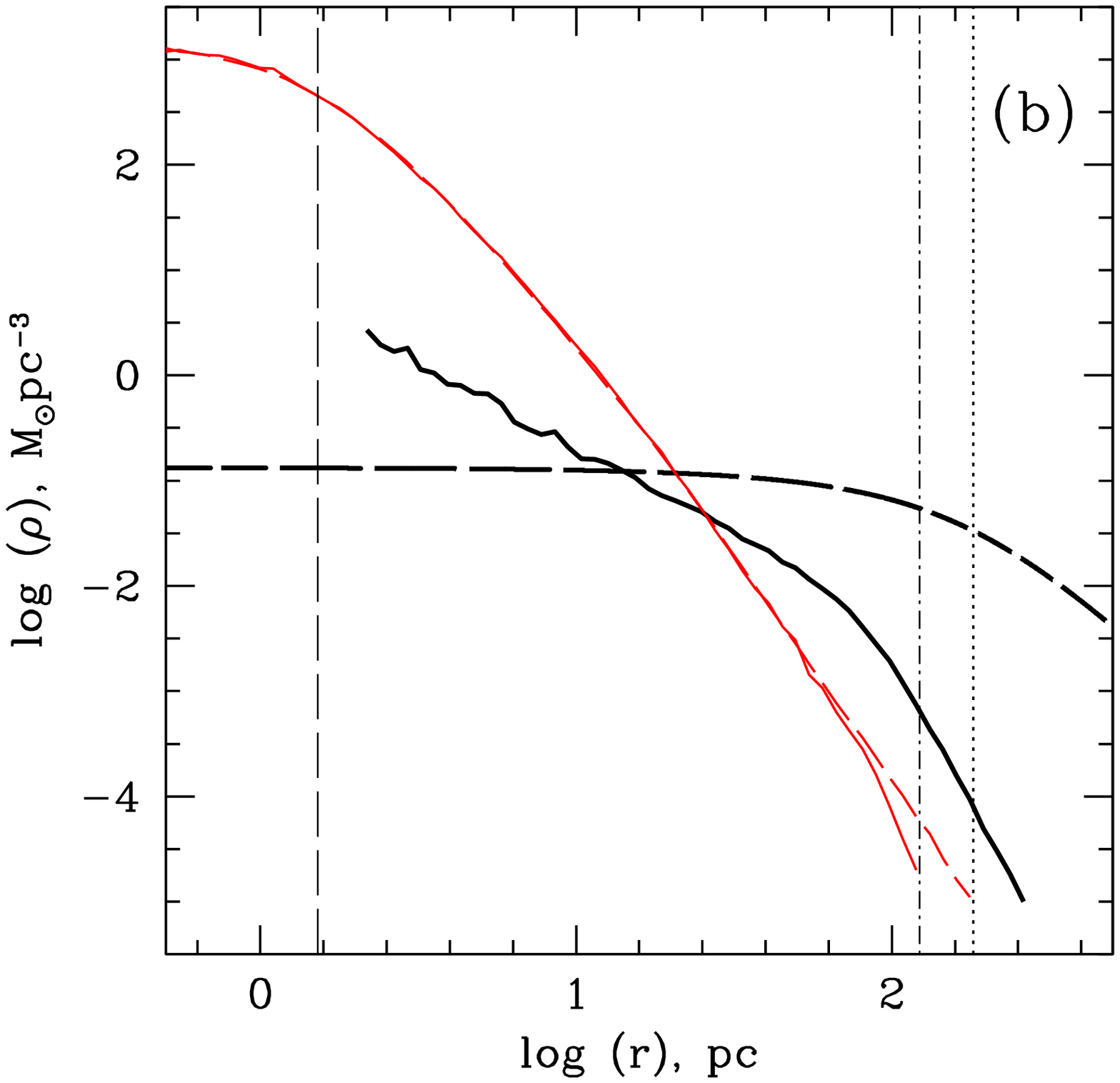}
\caption {Final radial density profiles for (a) NFW models and (b) Burkert models
(for the time interval $t=2.67\dots 3$~Gyr). Thick and thin (colored red in
electronic edition) lines correspond to DM and stars, respectively. Long-dashed
lines show the profiles of DM in the absence of stars and tidal field
(analytical models) and stars in the absence of DM (model S). Solid lines
correspond to SDO$_{n,b}$ models. Vertical long-dashed, short-dashed,
dash-dotted, and dotted lines mark the values of $\epsilon_{\rm DM}$, $r_m$,
current tidal radius for SDO models $r_{\rm tid}$, and $r_s$.
\label{rho2} }
\end{figure*}

\begin{figure*}
\plottwo{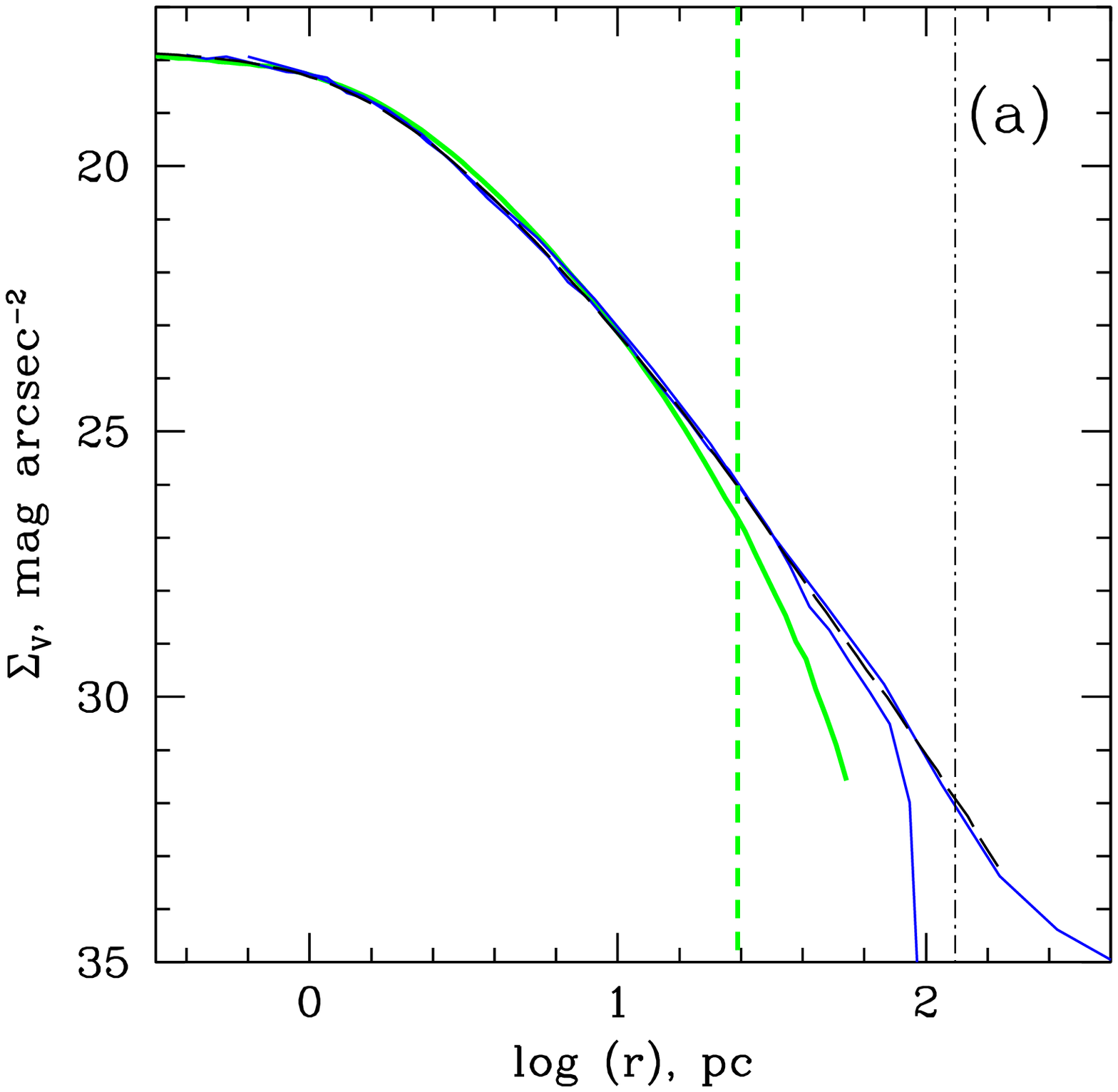}{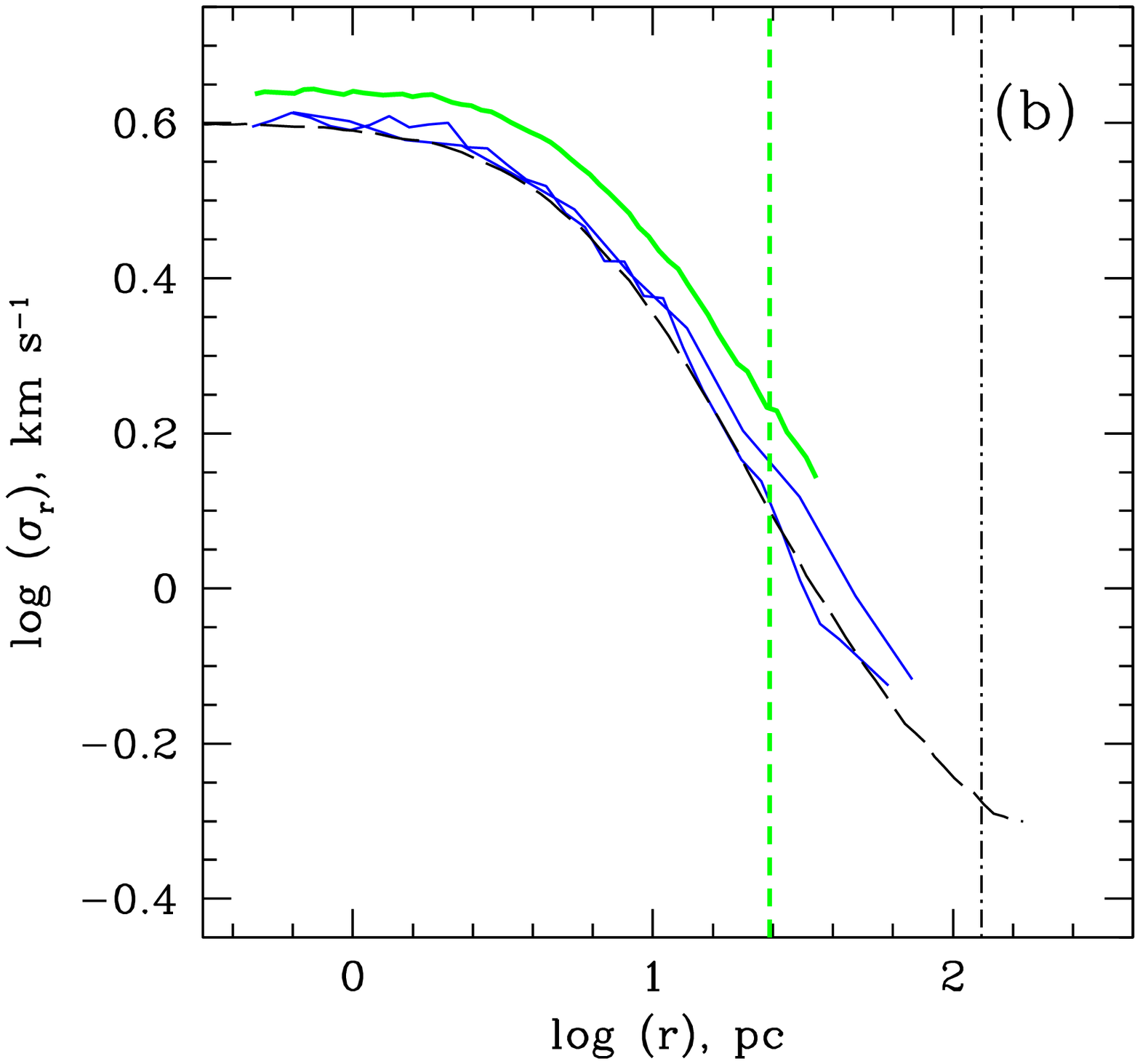}
\caption {Final radial profiles of observable quantities (for the time interval 
$t=2.67\dots 3$~Gyr -- the same as in Fig.~\ref{rho2}). (a) V-band surface
brightness $\Sigma_V$. (b) Line-of-sight stellar velocity dispersion
$\sigma_r$. Thick and thin solid lines (colored green and blue in electronic
edition, respectively) correspond to models SDO$_n$ and SDO$_b$,
respectively. (There are two DSO$_b$ profiles: the upper one
corresponds to stars along the major axis of the cluster, and the lower one is
for stars in the plane perpendicular to the major axis.)  Long-dashed lines show
the profiles for the model S.  Thick vertical short-dashed line marks the value
of $r_m$ for model SDO$_n$ (see Table~\ref{tab2}). Vertical dash-dotted lines
correspond to the current tidal radius for SDO models $r_{\rm tid}$. To make
these plots we use the projection technique described in Appendix~B of Paper~I.
\label{sigma} }
\end{figure*}


In Figure~\ref{DM_mass}a, we show the evolution of the gravitationally bound DM
mass for model DO$_{n}$ (solid thin line). As you can see, tidal stripping is
severe in this model, with only $\sim 3.8$\% of the total mass surviving as a
bound structure after 11 orbits (see Table~\ref{tab2}). This behavior is
similar to the critical case of \citet{hay03} with $r_{\rm tid}/r_s=2.1$ (see
their Fig.~7), which separates their models staying relatively intact (with
$r_{\rm tid}/r_s>2.1$) and completely disrupted models (with $r_{\rm
tid}/r_s<2.1$).  In our case, this ratio is slightly smaller than 2.1: $r_{\rm
tid}/r_s=1.73$. Slightly larger resilience to tidal disruption exhibited by
our model DO$_{n}$ can be explained by the fact that \citet{hay03} used ``local
Maxwellian approximation'' to set up the initial particle distribution in their
models, which according to \citet{kaz04} leads to artificially high disruption
rate for NFW subhalos.

Even for the very last snapshot of model DO$_{n}$ (at $t=3$~Gyr), the
stripped-down subhalo appears to be reasonably stable against total disruption
by the tidal forces. Indeed, the binding radius of the remnant $r_{\rm
bind}=15$~pc is significantly smaller than the current value of the tidal radius
$r_{\rm tid}=56$~pc even at the very end of the simulations. (To estimate
$r_{\rm tid}$, we use the actual value of the pericentric distance of
$R_p=413$~pc which is smaller than the original value of $R_p=600$~pc because of
the dynamic friction experienced by the subhalo moving in the halo of unbound DM
particles.)  We expect thus that our NFW subhalo should survive as a
gravitationally bound remnant for a few more orbits (and perhaps indefinitely).

The fate of a Burkert DM subhalo orbiting inside a Burkert host galaxy (model
DO$_{b}$) is completely different, as can be seen in Figure~\ref{DM_mass}b
(solid thin line).  After only 3 pericentric passages, the subhalo ceases to
exist as a coherent structure. To make sure that the failure to find a bound
structure after $t=0.79$~Gyr is not an artifact of our halo finding algorithm,
we measure both binding radius $r_{\rm bind}=625$~pc and tidal radius $r_{\rm
tid}=175$~pc for the last snapshot containing a bound structure. As you can see,
at this point the subhalo is bound to be dispersed by tidal forces as its
current tidal radius is significantly (almost 4 times) smaller than the binding
radius.

Two main conclusions we arrive at in this section are (1) our NFW DM-only
subhalo is relatively resilient to total tidal disruption, and (2) a Burkert
DM subhalo orbiting inside a Burkert host galaxy is much easier to disrupt that for
the case of an NFW DM subhalo orbiting inside an NFW host.

\subsection{Tidal Stripping of Hybrid GCs}

We are now turning to the results of our two principle models: SDO$_n$ and
SDO$_b$. In these models, a hybrid GC (stars $+$ DM) is orbiting inside the same
host galaxies and along the same eccentric orbit as in models DO$_{n,b}$. Before
being placed in the static potential of the host galaxy, we allow stars and DM
to relax in isolation for 120~Myr (model SDO$_n$) and 170~Myr (model SDO$_b$).

As you can see in Figure~\ref{DM_mass}, presence of a stellar core, with a mass
of mere $\sim 0.9$\% of the total mass, inside a DM subhalo makes the subhalo
much more resilient to tidal forces disruption. In the case of NFW profile
(Figure~\ref{DM_mass}a; see also Table~\ref{tab2}), the DM mass of the tidally
stripped remnant at $t\simeq 3$~Gyr is 7 times larger in the presence of the
stellar core than in the DM-only case. It is also 3 times larger than the mass
of the stellar core. For Burkert profile, the difference is even more dramatic:
whereas a starless Burkert subhalo is completely disrupted after 3 orbits, the
subhalo with a stellar core survives till the end of the simulations, with $\sim
32$\% of the total mass of the remnant being in DM form.

Figure~\ref{E_tot} provides an explanation for such a marked difference. As we
discussed in \S~\ref{prelim}, both NFW and Burkert halos become unbound if
instantaneously truncated below a certain radius ($0.77 r_s$ for NFW profile,
and $1.66 r_s$ for Burkert profile). As you can see in Figure~\ref{E_tot}, the
presence of a stellar GC-like cluster at the center of either halo makes it
significantly more bound. In the case of NFW profile (thick solid line), the
halo becomes bound for virtually any truncation radius $r$ (being similar in
this regard to a singular isothermal sphere). For Burkert halo (thin solid
line), there is still a range in truncation radius $r$ where the halo is
formally unbound, but the maximum positive value of $E_{\rm tot}$ is
significantly lower in the presence of a stellar core. Moreover, for $r\lesssim
0.32 r_s$ the halo again becomes bound (see Figure~\ref{E_tot}).

After 7 orbits or $\sim 2$~Gyr, the mass of DM gravitationally bound to the
remnant in model SDO$_b$ becomes smaller than the mass of the stellar core. At
this point, the rate of the mass loss due to tidal stripping becomes smaller,
but still non-negligible (see Figure~\ref{DM_mass}b).  We tested the possibility
that the observed decrease in bound DM mass after $t\simeq 2$~Gyr is a numerical
artifact, caused by a small number ($<10^4$) of DM particles attached to the
remnant, by resimulating the late evolution of the subhalo in the absence of the
static gravitational potential of the host galaxy. For this, we used the
gravitationally bound remnant from the $t=2.13$~Gyr snapshot of model SDO$_b$.
The bound DM mass evolution for this additional run is shown with dotted line in
Figure~\ref{DM_mass}b. As you can see, in the absence of external tidal field
the bound DM mass of the remnant stays virtually constant. It appears thus very
unlikely that the late time DM stripping observed in model SDO$_b$ is caused by
numerical artifacts (such as evaporation of DM particles due to two-body
interactions).

In Figure~\ref{column} we show DM column density maps for models SDO$_n$ (left
panels) and SDO$_b$ (right panels), for two moments of time: after $\sim 1.5$
orbits (top panels) and after $10-11$ orbits (bottom panels). Two concentric
circles show the range of radial distances covered by the models. At the earlier
moment of time, you can see a long stream of DM particles, stripped during the
first pericentric passage, with a relatively massive bound structure in the
center of the stream. At the end of the simulations, multiple streams of
stripped particles mix together to create a fuzzy thick disk of
DM. The gravitationally bound remnant is barely visible at this point (especially
for Burkert model).

In Figure~\ref{DM_rho} we show averaged DM radial density profiles for
$t=1.61\dots 3$~Gyr for models D$_{n,b}$ (thick solid lines), SD$_{n,b}$ (thin
solid lines), and SDO$_{n,b}$ (short-dashed lines). As you can see, the DM
density in the innermost part of the NFW halo is not modified by external tidal
field.  Outside of the stars-dominated central region, the DM density profile
becomes significantly steeper in the presence of tidal field. At this point, in
models SDO$_{n,b}$ DM is stripped down to the original scale radius $r_s$.

Analysis of Table~\ref{tab2} shows that the parameters of our stellar clusters
at the end of the simulations are very similar for all our models containing
stars (S, SD$_{n,b}$, and SDO$_{n,b}$). For SD$_n$ and SDO$_n$ models, the
presence of relatively large amounts of DM at the center of the cluster leads to
somewhat larger values of the central dispersion $\sigma_c$ and smaller values
of the King core radius $r_0$. The value of the apparent mass-to-light ratio
$\Upsilon$, defined as (see Paper~I)

\begin{equation}
\Upsilon = \frac{9 \sigma_0^2 \Upsilon_{\rm GC}}{2\pi G \zeta_0 R_{hb}},
\label{eqUpsilon}
\end{equation}

\noindent where $\Upsilon_{\rm GC}\simeq 1.45$ is the assumed baryonic
mass-to-light ratio in GCs and $\zeta_0$ is the projected stellar surface mass density
at the center of the cluster, is larger by $\sim 18$\% for a stellar
cluster in NFW halo. Overall, the presence of an external tidal field does not seem
to change the global structural and dynamic parameters of GCs with DM in a
noticeable way.

As can be seen in Figure~\ref{rho}, tidal stripping does not modify
significantly the stellar radial density profiles of hybrid GCs: at any moment of
time, density profiles for SDO$_{n,b}$ models (dotted lines) are very close
to the density profiles of SD$_{n,b}$ models (solid lines). 



In Figure~\ref{rho2} we show the final radial density profiles for both stars
and DM for models SDO$_n$ and SDO$_b$. As you can see, DM still dominates stars
in the outskirts of the stellar cluster. In the stars-dominated area, the DM density
profile is steeper because of the adiabatic contraction of DM in the presence of
stars. DM density profiles are dramatically modified both by the presence of a
stellar core and by tidal stripping. 

In model SDO$_b$, stellar cluster becomes tidally limited by the end of the
simulations. By $t=3$~Gyr, around 0.6\% of stars (or $\sim 60$ stellar
particles) are not gravitationally bound to the cluster, forming distinctive
trailing and leading stellar tidal tails. (Conversely, not a single stellar
particle has been tidally stripped by $t=3$~Gyr in model SDO$_n$.) In model
SDO$_b$, the shape of the cluster becomes increasingly non-spherical at large
radii at $t\sim 3$~Gyr. As can be seen in Figure~\ref{sigma}a, there is an
excess of stars in the outskirts of the cluster along its major axis, and a
sharp tidal cutoff around the analytical tidal radius $r_{\rm tid}$ near the
plane perpendicular to the major axis. (To calculate surface brightness and
velocity dispersion profiles for SDO$_b$ model, we used all stellar particles --
both bound and unbound.)

As Figure~\ref{sigma}a shows, the final surface brightness profiles of the
stellar clusters in SDO$_{n,b}$ models look remarkably similar to the
corresponding profiles of Galactic GCs. (We assumed that the baryonic V-band
mass-to-light ratio in GCs is $\Upsilon_{\rm GC}=1.45$ -- see discussion in
Paper~I.) The ``dent'' feature observed in the surface brightness profile of a
freshly relaxed hybrid (DM $+$ stars) GC (see Paper~I) has been removed in
models SDO$_{n,b}$ by secular evolution of the stellar cluster (see
\S~\ref{secular}). The apparent ``tidal'' cutoff in the outer surface brightness
profile of model SDO$_{n}$ is caused by the presence of significant amounts of
DM in the outskirts of the cluster (similarly to Paper~I). In the case of
Burkert halo (model SDO$_{b}$), the cutoff is of truly tidal nature.

In Figure~\ref{sigma}b we show the final line-of-sight velocity dispersion
profiles for the stellar clusters in SDO$_{n,b}$ models. In the case of NFW halo
(thick solid line), the line-of-sight velocity dispersion appears to be
uniformly inflated by a small factor ($\sim 10$\%) across all radii, which can
be misinterpreted as a purely stellar cluster with a somewhat larger value of
baryonic mass-to-light ratio. For the Burkert halo (thin solid lines), the
radial velocity dispersion profile for the stars near the plane perpendicular to
the major axis is very close to the profile of a purely stellar cluster, whereas
the stars along the major axis show signs of being tidally heated.

\section{CONCLUSIONS}
\label{conclusions}

DM subhalos with Burkert density profile are much easier to disrupt tidally in
the potential of the host galaxy than subhalos with NFW profile. We link this
effect to the difference in the binding properties of both types of halos, with
the total energy of the Burkert halos becoming positive if truncated at a
significantly larger radius than for NFW halos.  Setting a low-mass ($\sim 1$\%
of the total mass) dense stellar core at the center of either NFW or Burkert
halo makes them much more resilient to tidal disruption.

Primordial GCs with NFW DM halo can survive severe tidal stripping in the host
galaxy, with DM still being the dominant mass component (though not by a large
margin) in the tidally stripped-down remnant.  DM is concentrated in the
outskirts of the remnant. As a result, an apparent ``tidal'' cutoff in the
radial surface brightness profile in isolated warm collapse models (Paper~I),
caused by the presence of DM, is also present in our tidally stripped NFW
model. 

We used warm collapse hybrid models to show that neither secular evolution of the
stellar cluster nor severe tidal stripping change noticeably the inferred
core mass-to-light ratio $\Upsilon$. For both flat-core and cuspy DM halo
profiles, $\Upsilon$ stays close to the purely baryonic value.

Tidal stripping can remove almost all DM from primordial GCs with Burkert DM
halo. The remaining DM is dynamically unimportant, and cannot prevent stars
being stripped off by tidal forces of the host galaxy. This result makes GCs
possessing obvious tidal tails (the only known example being Palomar~5) be fully
consistent with primordial scenarios of GC formation.

Secular evolution of a DM-dominated GC does not change the main results of
Paper~I derived for freshly relaxed warm-collapse systems. In particular, in
both unevolved and evolved clusters, presence of significant amounts of DM in
the outskirts of the cluster manifest itself as a ``tidal'' cutoff in the outer
part of the radial surface brightness profile. It is not clear if the
``extratidal'' features of cold collapse models from Paper~I will be preserved
in significantly evolved clusters. As we discussed in
\S~\ref{physical}, simulating a cold-collapse hybrid GC to address this issue is
not feasible with the present day technology.

The above results reinforce our conclusion from Paper~I that a presence of
obvious tidal tails is probably the only observational evidence which can
reliably rule out a presence of significant amounts of DM in GCs. (But it cannot
rule out primordial scenarios of GC formation.)

To summarize, the results presented in both Paper~I and this paper suggest that
the whole range of features seen in Milky Way GCs (from apparently truncated
profiles of some clusters to the extended tidal tails of Palomar~5) can be
consistent with the primordial picture of GC formation, given that there was a
range of the initial virial ratios for stellar clusters (from ``cold'' to
``hot'' collapses), tidal stripping histories, and/or inner DM density profiles.

\acknowledgements We would like to thank Volker Springel for his help with introducing
an external static potential in GADGET. S. M. is partially supported by
SHARCNet. The simulations reported in this paper were carried out on McKenzie
cluster at the Canadian Institute for Theoretical Astrophysics.


\end{document}